\documentclass[useAMS,usenatbib]{mn2e}

\usepackage{natbib, aas_macros}
\usepackage{ulem}
\usepackage[dvips]{graphicx}
\usepackage{wrapfig}
\graphicspath{{./figs/}}
\usepackage{color}
\usepackage{amsmath}
\usepackage{amssymb}

\newcommand{\Msun}{{\rm M_{\odot}}}
\newcommand{\hMsun}{h^{-1} {\rm M_{\odot}}}

\newcommand{\Zsun}{{\rm Z_{\odot}}}
\newcommand{\fesc}{f_{\rm esc}}

\newcommand{\SFR}{{\rm SFR}}
\newcommand{\Lsun}{{\rm L_{\odot}}}

\newcommand{\art}{\rm ART^{2}}
\newcommand{\A}{\rm \AA}

\newcommand{\hMpc}{h^{-1} \rm Mpc}

\newcommand{\Mstar}{\rm M_{\rm{star}}}

\newcommand{\lya}{\rm {Ly{\alpha}}}

\newcommand{\Mh}{M_{h}}
\newcommand{\Msunyr}{\rm M_{\odot}~ yr^{-1}}
\newcommand{\fescion}{f_{\rm esc}^{\rm ion}}

\newcommand{\Mst}{{ M_{\rm star}}}
\newcommand{\Mgas}{{ M_{\rm gas}}}
\newcommand{\Mdust}{{ M_{\rm dust}}}
\newcommand{\muv}{m_{\rm UV}}

\def\gtorder{\mathrel{\raise.3ex\hbox{$>$}\mkern-14mu
    \lower0.6ex\hbox{$\sim$}}}
\def\ltorder{\mathrel{\raise.3ex\hbox{$<$}\mkern-14mu
    \lower0.6ex\hbox{$\sim$}}}

\title[Observational Properties of Simulated Galaxies at $z\simeq 6-12$]
{Observational Properties of Simulated Galaxies in Overdense and Average Regions at Redshifts 
$z\simeq 6-12$}
\author[Yajima et al.]
{Hidenobu Yajima$^{1, 2, 3}$\thanks{E-mail: yajima@roe.ac.uk},
Isaac Shlosman$^{3, 4}$, Emilio Romano-D\'{\i}az$^{5}$, Kentaro Nagamine$^{3, 6}$ 
\\
$^{1}$ Frontier Research Institute for Interdisciplinary Sciences, Tohoku University, Sendai 980-8578, Japan\\
$^{2}$ Astronomical Institute, Tohoku University, Sendai 980-8578, Japan\\
$^{3}$ Department of Earth \& Space Science, Graduate School of Science, Osaka University, 1-1 Machikaneyama, 
   Toyonaka, Osaka 560-0043, Japan\\
$^{4}$ Department of Physics \& Astronomy, University of Kentucky, Lexington, KY 40506-0055, USA\\
$^{5}$ Argelander Institut fuer Astronomie, University of Bonn, Auf dem Haegel, 71, D-53121 Bonn, Germany\\
$^{6}$ Department of Physics \& Astronomy, University of Nevada, Las Vegas, 4505 S. Maryland Pkwy, Las Vegas, NV 89154-4002, USA\\
}
 
\begin{document}

\date{Accepted ?; Received ??; in original form ???}

\pagerange{\pageref{firstpage}--\pageref{lastpage}} \pubyear{2008}

\maketitle

\label{firstpage}

%
%
\begin{abstract}
We use high-resolution zoom-in cosmological simulations of galaxies of Romano-D\'{i}az et al., 
post-processing them with a panchromatic three-dimensional radiation transfer code 
to obtain the galaxy UV luminosity function (LF) at $z\simeq 6-12$.
The galaxies are followed in a rare, heavily overdense region 
within a $\sim 5\sigma$ density peak, which can host high-$z$ quasars, and in an average density region, 
down to the stellar mass of $M_{\rm star}\sim 4\times 10^7\,\Msun$. 
We find that the overdense regions evolve at a substantially accelerated pace --- the most massive 
galaxy has grown to $M_{\rm star}\sim 8.4\times 10^{10}\,\Msun$ by $z = 6.3$, 
contains dust of $M_{\rm dust}\sim 4.1\times 10^8\,\Msun$, and  
is associated with a very high star formation rate, SFR\,$\sim 745\,\Msunyr$.
The attained SFR$ - M_{\rm star}$ correlation results in the {\it specific} SFR slowly increasing 
with $M_{\rm star}$. Most of the UV radiation in massive galaxies is absorbed by 
the dust, its escape fraction $\fesc$ is low, increasing slowly with time. 
Galaxies in the average region have less dust, and agree with the observed UV LF. 
The LF of the overdense region is substantially higher, and contains much brighter galaxies.   
The massive galaxies are bright in the infrared (IR) due to the dust thermal emission, with 
$L_{\rm IR}\sim 3.7\times 10^{12}\,\Lsun$ at $z = 6.3$, while  $L_{\rm IR} < 10^{11}\,\Lsun$ for the 
low-mass galaxies. Therefore, ALMA can probe massive galaxies in the 
overdense region up to $z\sim 10$ with a reasonable integration time. 
The UV spectral properties of disky galaxies depend significantly upon the viewing angle.
The stellar and dust masses of the most massive galaxy in the overdense region are comparable 
to those of the sub-millimetre galaxy (SMG) found by Riechers et al. at $z = 6.3$, 
while the modelled SFR and the sub-millimetre flux fall slightly below the observed one. 
Statistical significance of these similarities and differences will only become clear with the 
upcoming ALMA observations.
\end{abstract}

%
%

%
%

\section{Introduction}
Understanding galaxy evolution in the high-redshift universe is one of the major goals in contemporary astronomy. 
To reveal the mechanism(s) of this evolution, it is crucial to study the galactic spectra reflecting their physical 
state, e.g., star formation rate (SFR), stellar mass, metallicity, dust contents, and their spatial distributions. 
In this work,  we investigate the UV spectral properties (including the UV luminosity function) of high-$z$ galaxies 
in a rare overdense region of the 
universe in the wide redshift range of $z\simeq 6-12$, and compare them with an average density region. We calculate
the UV continuum, both intrinsic and attenuated by dust, estimate  the FIR emission from the dust, and analyze
their observational corollaries. 

At high redshifts, $z\gtrsim 6$, thousands of galaxies have been detected using the drop-out technique in the UV band 
--- the so-called Lyman break galaxies \citep[LBGs:~][]{Shapley03, Bouwens09, Bouwens10, Bouwens11b, Bouwens12, 
Bouwens14, Ouchi09b, McLure13, Ellis13, Oesch12, Oesch13, Finkelstein14}. These LBGs are typical star-forming 
galaxies at high-$z$, 
and have been used to probe the cosmic star formation (SF) history \citep[e.g.,][see also the review by Shlosman 
2013]{Madau99, Madau14, Steidel99, Ouchi04, Hopkins06SFR, Bouwens07, Bouwens14}. Another population of 
high-redshift starforming galaxies is the Lyman-$\alpha$ emitters (LAEs) which are identified by $\lya$ 
lines \citep[e.g.,][]{Iye06, Gronwall07, Ouchi10, Blanc11,  Vanzella11, Ono12, Shibuya12,  Finkelstein13}.
LBGs and LAEs are thought to be young starforming galaxies with little dust. 

Large column densities of dust attenuate the UV continuum and the $\lya$ photons, and galaxies become fainter than 
the detection limits for current observations. 
Massive galaxies can possess dust produced by type II supernovae (SNe) via earlier SF
\footnote{  
Note that the dominant channel for dust formation is being debated. 
For example, dust can also be grown in molecular clouds and produced in stellar winds from the AGB stars 
\citep[e.g.,][]{Draine03}. Since we focus on the early universe in this work, we assume that the dust is 
produced mainly by the SNe. 
}
\citep[e.g.,][]{Todini01, Schneider04, Nozawa07, Hirashita14}. 
This dust can absorb the UV photons and re-emit the energy in the infrared band as a thermal emission. 
Modern sub-millimetre telescopes, e.g., ALMA or the Herschel satellite, are gradually 
opening the window for detection of distant SMGs. A recently discovered  
SMG at $z=6.3$ \citep{Riechers13} exhibits $\SFR \sim 2900\,\Msunyr$ and a dust mass of $\Mdust \sim 
10^9\,\Msun$. Thus, at $z\gtorder 6$, the radiation properties of galaxies show a wide range, involving SMGs, LBGs, 
LAEs, etc. 

A number of cosmological simulations have been carried out to reproduce the typical starforming galaxies at high-$z$ in 
the mean 
density fields \citep[e.g.,][]{Nagamine04e, Romano-Diaz09, Finlator11, Yajima12b, Yajima12c, Yajima12f, Yajima14b, 
Jaacks13, Wise12a, Wise14, Paardekooper13}, or massive galaxies in rare overdense regions by using large computational 
boxes \citep[e.g.,][]{DiMatteo05, Li07, Cen12} or Constrained Realizations \citep[e.g.,][]{Romano-Diaz11b, Romano-Diaz14}. 
While the physical state of galaxies in various fields have been investigated, their spectral properties are not
well known, requiring detailed radiative transfer (RT) calculations. Simulations, which attempt to infer observational 
properties without RT, must make additional assumptions, 
such as the spatial distribution of dust, etc. \citep[e.g.,][]{Nagamine10b, Dayal12, Shimizu14}.

As a next step,  the dust extinction should be estimated by means of the RT, 
and the emergent radiation properties can be compared to the observations. 
By applying RT, the actual stellar and dust distributions will be  considered, and the optical depth for relevant
photons will be calculated. 
\citet{Dayal13} investigated radiation properties of galaxies in cosmological simulations by using a simple dust 
screening model, and reproduced the statistical properties of observed LBGs at $z\ge 6$. However, 
since the dust is distributed inhomogeneously in galaxies,
numerical simulations are required to calculate the RT along various directions and from a large  
number of stars. Combining cosmological Smoothed Particle Hydrodynamics (SPH) simulations with the post-processed 
RT, \citet{Yajima14b} analyzed the dust extinction of starforming galaxies at $z=3$,
and have reproduced the observed luminosity function (LF) adopting a SN dust model. 
The LF was found to be moderately insensitive to dust properties, e.g.,  composition, grain size, and 
dust-to-metal mass ratio, despite that the colour in the UV bands is known to depend sensitively on the dust 
properties. In particular, they found that the dust composition cannot be of a purely silicate type. Instead,
the graphite type should be the dominant component in order to reproduce the observed extinction level and the 
rest-frame UV LF of LBGs at this redshift. 

Using cosmological adaptive mesh refinement (AMR) simulations and RT calculations, \citet{Kimm13} 
have recently analyzed the dust extinction of the UV stellar radiation in galaxies at $z=7$ with stellar masses of 
$5\times 10^8\,\Msun - 2.5\times 10^{10}\,\Msun$, and have shown that most of the UV 
radiation is absorbed by dust. In a companion paper, \citet{Cen14} have further investigated the infrared 
properties of these model galaxies.
These calculations have followed a density peak in a box of $120~h^{-3}\rm Mpc^3$, 
corresponding to a $1.8 \sigma$ fluctuation, which led to a peak SFR\,$\sim 100\,\Msunyr$ at $z\sim 7$, and 
formation of a galaxy cluster of $\sim 3\times 10^{14}\,\Msun$ by $z\sim 0$.
The sample was limited to $z=7$ and to 198 massive galaxies in the biased region --- the observational properties of 
galaxies did not include the rare overdense regions analysed here which can serve as an environment for the
high-$z$ quasars. 

Here we investigate the UV and IR observational properties of high-$z$ galaxies in rare overdense and average
regions, using the cosmological simulations of \citet{Romano-Diaz14} \citep[see also][]{Romano-Diaz11b}. 
They have shown that massive galaxies tend to have extended gas disks due to the efficient gas accretion from  
the intergalactic medium (IGM) \citep[see also][]{Pawlik11}.
The presence of extended disks in these  galaxies 
appear robust against the strong radiative feedback \citep{Pawlik13}.
Hence, we anticipate that when the gaseous/stellar disks form, their spectral properties change significantly, 
depending on the viewing angles, i.e., the ``orientation effect'' \citep{Yajima12c, Verhamme12}. 
These works have analyzed the orientation effect for low-mass starforming galaxies 
$\Mh\sim 10^{11}\,\Msun$ like LAEs. 
For the first time, we turn our attention to the orientation effects in the UV properties of high-$z$ massive 
galaxies in a cosmological context. 

Our paper is organized as follows.
We describe the simulations and our method for the radiative transfer calculations in Section~\ref{sec:model}. In 
Section~\ref{sec:result}, 
we present our results, show the simulated UV fluxes, escape fraction of non-ionizing photons, the infrared dust 
emission 
with detectability by ALMA, as well as investigate the effect of the disk galaxy orientation on the UV properties. 
In Section~\ref{sec:discussion}, we discuss our results and summarize our main conclusions.

%
%

\section{Numerical Modelling}
\label{sec:model}

\subsection{Hydrodynamic Simulations}

We use the high-resolution numerical simulations of \citet{Romano-Diaz11b, Romano-Diaz14}, 
which were performed with the modified tree-particle-mesh SPH code 
GADGET-3 \citep[originally described in][]{Springel05e} 
in its conservative entropy formulation \citep{Springel02}.
The simulations were evolved from $z = 199$ to $z\sim 6$. 
They include radiative cooling by H,
He and metals \citep[e.g.,][]{Choi09}, star formation, stellar feedback, a galactic
wind model, 
and a sub-resolution model for multiphase interstellar medium \citep[ISM;][]{Springel03a}. 
In this model, starforming gas particles contain both the cold phase (which contributes to the
gas mass and forms stars) and the
hot phase (that results from the SN heating and determines the gas pressure). 

The SF prescription is based on the
{\it Pressure} model \citep{Schaye08, Choi10} which reduces the high-$z$ SFR 
relative to the prescription by \citet{Springel03b}. 
The {\it Pressure} model uses the conversion between the gas Jeans column density, 
$\Sigma_{\rm gas,J}$,  and 
its mass density, i.e., 
$\Sigma_{\rm gas,J}\sim \rho_{\rm gas}L_{\rm J} = \sqrt{(\gamma/G)f_{\rm g}P_{\rm tot}}$ 
\citep{Schaye08,Choi10}, where $L_{\rm J}$ is the Jeans length, $f_{\rm g}$ is the gas mass fraction, 
$P_{\rm tot}$ its total pressure, i.e., thermal, turbulent, etc., and $\gamma=5/3$. 
If the gas column density $\Sigma_{\rm gas}$ exceeds $\sim 10\,{\rm \Msun\,pc^{-2}}$, 
the stars form with the rate $\Sigma_{\rm SFR} = 2.5 (\Sigma_{\rm gas} / 1\,{\rm \Msun\,pc^{-2}})^{1.4} 
{\rm \Msun\,yr^{-1}\,kpc^{-2}}$ \citep{Kennicutt98, Kennicutt98b}.
Within this model, the SF only takes place 
when the gas density rises above the threshold, $n_{\rm H,SF} = 0.6\,{\rm cm^{-3}}$.
The zoom-in technique was employed in the simulations, with the gas being
present only in the high-resolution region.  Since the UV background intensity is not strong 
until the end of the cosmic reionization epoch \citep[e.g.,][]{Faucher-Giguere09}, we do not consider it.
Note that, however, the UV background can locally be strong at near star-forming galaxies and has impact on galaxy evolution
\citep[e.g.,][]{Wise12a, Wise14}.

Low-mass galaxies are expected to be sensitive to stellar feedback, especially in the early universe, because of their 
shallow potential well, making it easier for the gas to be pushed out. In an associated work, we analyze 
various prescriptions for feedback and compare the results with observations \citep[][]{Sadoun15}. 
Current runs are based on the \citet{Springel03a} phenomenological wind model. This wind is  
driven by the SNe heating of the hot phase in the ISM whose cooling timescale is long. The cold phase of 
the ISM is heated and evaporated via thermal conduction from the hot phase. The mass-transfer equations 
between the phases are solved, based on the equilibrium model of \citet{Springel03a}. The `wind' SPH 
particles are temporarily not subject to hydrodynamical forces. 
The timescale for turning off the hydrodynamic interactions for the wind particle is either 50\,Myr or 
when the background gas density has fallen by a factor of 10, whichever happens first.

\begin{table*}
\begin{center}
{Parameters of Cosmological Simulations}
\begin{tabular}{ccccccccc}
\hline  & $N_{\rm SPH}$ & $R_{\rm box}$ & $R_{\rm box}^{\rm zoom}$ &$m_{\rm DM}$ & $m_{\rm SPH}$ & $m_{\rm
star}$ &$\epsilon$  \\  & & ($h^{-1} \rm Mpc$) & ($h^{-1} \rm Mpc$) & $(h^{-1} \Msun)$ & $(h^{-1} \Msun)$ &
$(h^{-1} \Msun)$ &$(\rm pc, comoving)$ &  \\
\hline
CR     & $1024^{3}$&  20  &3.5& $4.66 \times 10^{5}$ & $1.11 \times 10^{5}$ & $5.55 \times 10^{4}$ & 300  \\
UCR   & $512^{3}$ & 20 & 7.0 &$3.73 \times 10^{6}$   & $8.90 \times 10^{5}$ & $4.40 \times 10^{5}$ & 300  \\
\hline
\end{tabular}
\caption{
$N_{\rm SPH}$ is an effective number of the SPH particles in the zoom-in regions.
$R_{\rm box}$ and $R_{\rm box}^{\rm zoom}$ are radii of the entire computational box and the zoom-in regions.
$m_{\rm DM}$,  $m_{\rm SPH}$ and $m_{\rm star}$ are the masses of DM, gas and stellar particles.
$\epsilon$ is the gravitational softening length (comoving).
}
\label{tab:param}
\end{center}\end{table*}

\subsection{Simulation Setup and Initial Conditions}

\begin{figure}
\begin{center}
\includegraphics[scale=0.5]{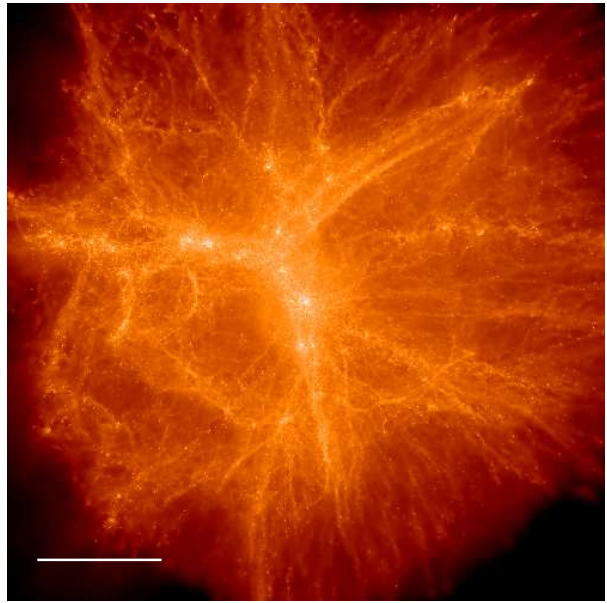}
\caption{
Projected gas density of the CR run at $z = 6.3$. The white bar represents $1\,h^{-1}{\rm Mpc}$ in comoving 
coordinates.
}
\label{fig:whole}
\end{center}
\end{figure}

\begin{figure*}
\begin{center}
\includegraphics[scale=0.8]{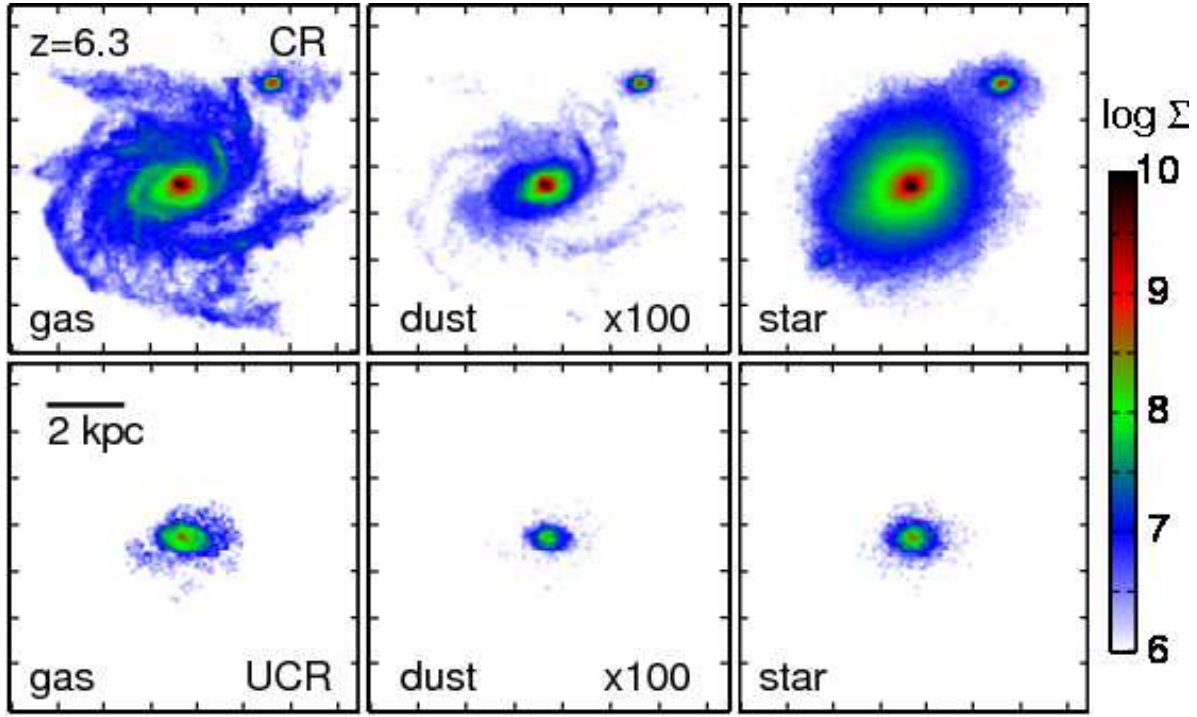}
\caption{
Projected gas, dust and stellar densities at $z = 6.3$. The colour palette represents the mass density in units 
of $\Msun\,{\rm kpc^{-2}}$ on the log scale.
The upper row represents the most massive galaxy in the CR run ($\Mst\sim 8.4\times 10^{10}\,\Msun$ and
$\Mgas\sim 4.8\times 10^{10}\,\Msun$). The lower row shows the most massive galaxy in the UCR run, $\Mst
\sim 1.5\times 10^9\,\Msun$ and $\Mgas\sim 3.1\times 10^9\,\Msun$.
The dust densities have been artificially boosted by factor of 100 for easy comparison with others in this figure. 
Each panel size is 9\,kpc in physical units.
}
\label{fig:2dgas}
\end{center}
\end{figure*}

\citet{Romano-Diaz11b, Romano-Diaz14} have used two types
of initial conditions (ICs). First, they followed up the high-redshift galaxy evolution in the mean density 
region in the universe, their so-called unconstrained (UCR) run. Second, they have applied the Constrained
Realization (CR) method \citep[e.g.,][]{Hoffman91, vdW96, Romano-Diaz11a} to the 
UCR initial conditions to simulate the rare overdense region of $\sim 5\sigma$. These ICs have been
constrained to include a seed of a massive halo of $\sim 10^{12}\,\hMsun$, 
projected to collapse by $z\sim 6$ according to the top-hat model, hereafter the CR run.
  
A CR of a Gaussian field is defined as a random
realization of such field, and is constructed to obey a set
of imposed linear constraints on this field, using the method of \citet{Hoffman91} and \citet[][see Appendix for a 
comprehensive explanation of the method]{Romano-Diaz11a}, by providing the optimal algorithm
to create the CRs, which have been used for setting the specific ICs applied here. 
The large-scale properties in both ICs are the same.

The ICs have been imposed onto a grid of 1024 (CR) and 512 (UCR) cells per dimension in a cubic box of 
20\,$\hMpc$. The $\Lambda$CDM-WMAP5 cosmology has been adopted with $\Omega_{\rm m} =
0.28$, $\Omega_{\Lambda} = 0.72$, and $\Omega_{\rm b} = 0.045$  \citep{Dunkley09}.
Here, $h = 0.701$ is the Hubble constant in units of $100\,{\rm km\,s^{-1}\,Mpc^{-1}}$. The power spectrum 
has been normalized by the linear rms amplitude of mass fluctuations in 8\,${h^{-1}}$Mpc spheres extrapolated to 
$z=0$, $\sigma_8 = 0.817$. The vacuum boundary conditions have been used, and the simulations have been
performed in comoving coordinates.

The gas is present only in the high-resolution regions
of radius $3.5\,h^{-1}$\,Mpc for the CR run, and $7\,h^{-1}$\,Mpc for the UCR run.
The UCR volume is matched to contain the same amount of total matter as the CR run.
The general parameters of the simulations are summarized in Table\,\ref{tab:param}.
The spatial resolution ranges from physical 23\,$h^{-1}$\,pc at $z = 12$ to 43\,$h^{-1}$\,pc at $z = 6$.
Galaxies are identified with the HOP group-finder algorithm \citep{Eisenstein98} 
for densities exceeding $0.01 n_{\rm H,SF}$ following \citet{Romano-Diaz14}.  

To study the spectral properties of simulated galaxies at $z = 6.3$, 7, 8, 9, 10, 11 and 12, 
we selected galaxies with $M_{\rm star}\gtorder 4\times 10^7\,\Msun$,
which corresponds to $\gtorder 720$ and $\gtorder 90$ stellar particles for CR and UCR runs, respectively. 
Our samples consist of 2273 and 1789 galaxies for the CR and UCR run, respectively,  
totaling 4062 galaxies. 

Figure\,1 shows a rendering of the projected gas density of the central zoomed-in region in the CR run at $z = 
6.3$. The 
yardstick represents the scale of $1\,h^{-1}$\,Mpc. Note the filamentary structure and the brightest 
regions representing galaxies. The most massive galaxy, which is a disk galaxy, is located near the centre of the 
zoomed-in region, and grows predominantly via the cold gas accretion from the cosmic filaments, supplemented by 
minor mergers \citep{Romano-Diaz14}.

\subsection{Radiation Transfer Calculations}

For the post-processing, we carry out the RT calculations using the Monte Carlo method in the 
`All-wavelength Radiation Transfer with Adaptive Refinement Tree' ($\art$) code to study the multi-wavelength 
properties of the model galaxies. The detailed prescriptions of the code have been presented in \citet{Li08} 
and \cite{Yajima12b}. We solve the RT of $10^6$ photon packets for each galaxy, which have shown 
 good convergence in the emergent luminosity and $\fesc$ \citep{Yajima12b}. In 
this work, we focus on the stellar continuum radiation at $\lambda > 912\,{\rm \AA}$, in the rest frame. 
Photons at UV-to-optical band can be absorbed by the interstellar dust, then re-emitted as a thermal emission 
in the IR. 
The dust temperature and its thermal emission have been estimated assuming a radiative equilibrium \citep{Li08}.
Our simulations follow the interactions between the photons and the dust on the adaptive refinement grids 
constructed by using positions of the SPH particles. 
The spatial resolution of the grids corresponds to that of the original hydrodynamics simulations. 
At high redshifts ($z\gtorder 6$) the dust is mainly produced by type-II SNe 
\citep[e.g.,][]{Hirashita14}. 
The SN dust model by \citet{Todini01} is assumed here, and its opacity curve was shown in \citet{Li08}.
Moreover, \citet{Yajima14b} 
have shown that the observed UV radiation properties of high-$z$ galaxies are 
reproduced well in the cosmological simulations with the RT calculations using the SN dust model. 

In the local star-forming galaxies, the dust mass is proportional to the metal mass \citep{Draine07}.
Using the metallicity of each SPH particle, we determine the dust mass of each cell under the assumption of 
constant dust-to-metal mass ratio, $M_{\rm dust}/M_{\rm metal}=0.4$, i.e.,
$M_{\rm dust} = 0.008\,M_{\rm gas}\,(Z/{Z_\odot})$, where $Z_\odot = 0.02$. 
Based on the two-phase model of ISM in GADGET-3 \citep{Springel05e}, 
we also separate the gas in each cell into the cold clumpy gas and the hot surrounding ISM. The gas phases 
are balanced by the thermal pressure, and the photons travel in the clumpy dusty gas \citep{Li08}. 

Current cosmological simulations still cannot resolve the propagation of the SNe shocks. In
our subgrid wind model, the wind particles move with the speed of  $\sim 500\,\rm km\; s^{-1}$, and we expect that
internal shocks will heat the wind to corresponding  $\sim 10^7$\,K.  At such temperatures
and wind densities, the dust is destroyed by thermal sputtering process in less than the wind crossing time
\citep[e.g.,][]{Nozawa06}. Therefore, we ignore the dust in the wind altogether.

The intrinsic spectral stellar energy distributions (SEDs) have been calculated with the stellar synthesis code, 
Starburst99 \citep{Leitherer99}
with a Salpeter initial mass function \citep{Salpeter55}.  
We have calculated also the radiation properties of galaxies at 
different viewing angles with 50 bins ($N_{\theta} = 5$ and  $N_{\phi} = 10$), and analysed the orientation 
effects. 

We have compared our CR run with a downgraded resolution of it (from $2\times 1024^3$
to $2\times 512^3$), both drawn from the same ICs and found no significant differences
in galaxy morphologies and the distributions and fractions of the ISM \citep{Romano-Diaz11b,Romano-Diaz14,
Sadoun15}.
Our spatial resolution by the end of the simulations is $\sim 43 h^{-1}$\,pc in the physical scale, which allows 
us to trace the clumpy 
ISM on scales $\gtorder 100$\,pc.


%
%

\section{Results}
\label{sec:result}

\begin{table}
\begin{center}
{Fitting Parameters}
\begin{tabular}{ccccccc}
\hline
 $X$ & $ Y$ & $\alpha$ & $\beta$   \\
\hline
${\rm log} \; M_{\rm star}$     & ${\rm log \; SFR} $   &1.18& -10.00   \\
\hline
${\rm log} \; M_{\rm star}$     &${\rm log} \; \fesc$  &-0.34& 2.15   \\
\hline
${\rm log} \; M_{\rm star}$     & $m_{\rm UV}^{\rm int}$   & -2.69& 51.00   \\
\hline
${\rm log} \; M_{\rm star}$     & $m_{\rm UV}$   &-1.88& 46.13   \\
\hline
${\rm log} \; M_{\rm star}$     & ${\rm log} \; S_{1.1}$   & 0.90 & -9.89\\
\hline
${\rm log} \; M_{\rm star}$     & ${\rm log} \; L_{\rm IR}$   &1.12 & 0.42\\
\hline
${\rm log} \; M_{\rm star}$     & ${\rm log} \; M_{\rm dust}$   &1.01 &  -3.22\\
\hline
&$Y = \alpha X + \beta$ \\
\hline
\end{tabular}
\caption{
Fitting parameters for the relation between stellar mass and observational properties at $z=6.3$.
The fitting equation is $Y = \alpha X + \beta$ with free parameters $\alpha$ and $\beta$.
$m_{\rm UV}^{\rm int}$ and $m_{\rm UV}$ give the UV flux without and with dust extinction, respectively. 
$S_{1.1}$ is the flux at $1.1~\rm mm$ in the observer's frame. 
The fitting parameters have been derived using all samples over the range of redshifts.
}
\label{table:fit}
\end{center}
\end{table}

We present our results starting with the physical properties of the galaxy population in the CR and UCR
runs, then continue with the appearance of these galaxies in the UV and IR bands. 
We present their UV luminosity functions, redshift evolution, and the effect of galactic morphology on the spectral
properties of high-$z$ galaxies. 

\subsection{Physical properties of galaxies}

The most massive galaxy in the CR run exhibits a resilient morphology of a gas-rich disk galaxy with a stellar 
bar and outer spiral arms, over a substantial period of time, $z\sim 6-10.5$, as shown in Figure~\ref{fig:2dgas} 
for $z=6.3$ 
and discussed in \citet{Romano-Diaz11b}. The bar is persistent, but is difficult to observe after $z\sim 6.4$
due to an ongoing merger. This figure displays the projected 
gas, dust and stellar densities at $z=6.3$. Located in the highly overdense region, the galaxy grows 
predominantly by a cold gas accretion.
The galaxy experiences a SFR of $\sim 745\,\Msunyr$
accumulating a $M_{\rm star}\sim 8.7\times 10^{10}\,\Msun$. 
The amount of dust is $\sim 4.1\times 10^8\,\Msun$, and the mean gas metallicity has reached $Z\sim 0.7\,\Zsun$. 
This dust is 
concentrated around the galactic center, while the stellar distribution is more extended and exhibits 
also a spheroidal component. 

In comparison, the most massive galaxy in the UCR run is more compact and spheroidal. 
The half-mass radius of the gaseous (stellar) distribution in the most massive UCR galaxy at $z\sim 6.3$ is about 
375\,pc (255\,pc). 
Most of the 
baryons are enclosed within the central kpc, and its SFR is $\sim 8.8\,\Msunyr$, which is a factor of
$\sim 80$ lower than that of the most massive object in the CR run. More detailed analysis of the
galactic morphology will be shown elsewhere. 

\begin{figure}
\includegraphics[scale=0.57]{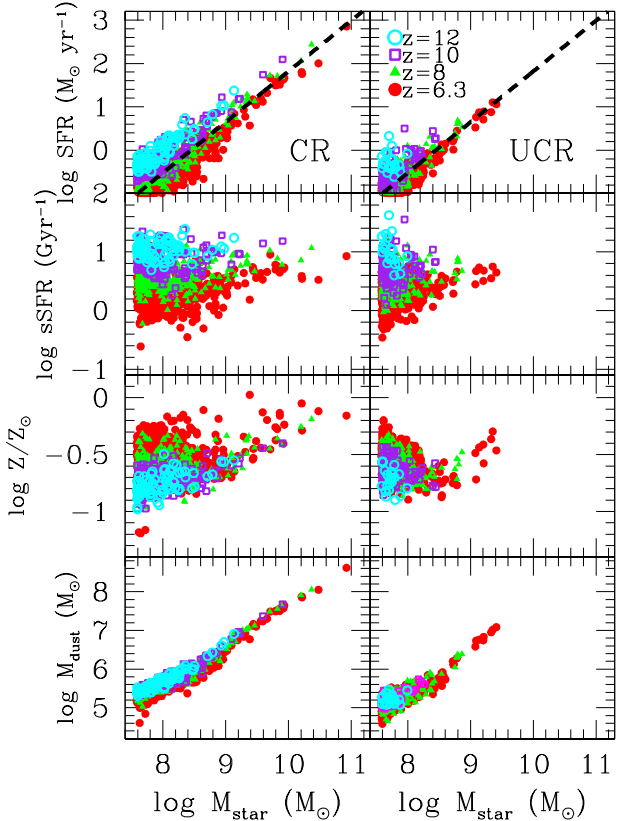}
\caption{
Star formation rate (SFR), specific star formation rate (sSFR), mean gas metallicity in units of the solar 
metallicity ($Z/Z_\odot$), and galactic dust mass ($M_{\rm dust}$), as a 
function of the galaxy stellar mass ($M_{\rm star}$). 
The different  symbols represent the different redshifts, $z = 6.3, 8, 10$ and 12.
}
\label{fig:sfr}
\end{figure}

The evolution of SFR, gas metallicity and dust mass as a function of the galaxy 
stellar mass are shown in Figure~\ref{fig:sfr}. 
For each redshift, the SFR increases nearly linearly with the stellar mass and its slope $\alpha$ 
in the log-log plot ranges within $\sim 1.00-1.27$, which is consistent with the previous works 
\citep[e.g.,][]{Finlator11}.
The slopes for all the samples are summarized in Table~\ref{table:fit}.
Such slopes imply a weak dependence of the 
specific SFR (${\rm sSFR}$) on $M_{\rm star}$, where sSFR$\equiv \SFR/{\Mst}$. 
This trend is also supported by observations \citep[e.g.,][]{Ono10A}.
As $\alpha\gtorder 1$, the sSFR slightly increases 
with $M_{\rm star}$ --- contrary to what has been found for lower redshifts \citep{Stark09, Stark13}. 
We find also that the sSFR decreases somewhat 
with redshift, as can be seen in the second panels (from the top) of Figure~\ref{fig:sfr}.
This trend and values are consistent with recent simulations \citep{Nagamine10b,Dayal13, Biffi13} 
and observations \citep{Stark13, Oesch14}.
Such a trend seems to reflect dependence of the sSFR on the gas fraction 
in galaxies which decreases with redshift for the low-mass galaxies. Some of the lower-mass galaxies are 
affected more by the stellar feedback, while the higher-mass 
galaxies retain much of their gas, keeping their high SFRs. 
This causes a substantially higher dispersion in the SFR for galaxies with $M_{\rm star}\ltorder 
10^9\,\Msun$. 

We do not observe any clear difference in the SFR -- $\Mst$ relationship between the CR and UCR runs, except the
presence of much higher galaxy masses in the former.
The SFR of most galaxies in the CR and UCR runs is fueled by the cold gas accretion rather than by major mergers, 
as shown by \citet{Romano-Diaz14}.  Obviously, the upper cutoff of the SFR lies much lower for the UCR run
because of the absence of massive galaxies in the normal region at $z\sim 6$. This cutoff is
increasing with time, e.g., it is $\sim 124\,\Msunyr$ at $z\sim 10$ and $745\,\Msunyr$ at $z\sim 6.3$, for the
CR run. 

One expects that galaxy evolution process is greatly accelerated in the overdense regions, which is supported by 
the evolution of the CR galaxies.
The third panels (from the top) of Figure~\ref{fig:sfr} show 
that the gas metallicity $Z/\Zsun$ has a large dispersion,
because the gas in low-mass galaxies is easily blown away by the SN feedback. 
In these objects, the metallicity can quickly rise to log\,($Z/\Zsun{\rm )}\sim -0.5$ once the stars form, 
and it will lead to a large dispersion, reflecting the variation in the SFRs. 
On the other hand, massive galaxies show a tighter correlation owing to the steady gas supply from the IGM 
filaments and minor mergers --- the width of quartiles to median values at each mass bin decreases with increasing 
mass. Metallicity of some massive galaxies in the CR run reaches a nearly solar abundance 
at $z\sim 6.3$. Already at $z\sim 10$, the most massive galaxy exhibits log\,($Z/\Zsun{\rm )}\sim -0.4$ due to 
early SF. 
There is little difference in the value and the spread between the CR and UCR runs. Most of the difference
again becomes visible for $M_{\rm star}\gtorder 10^9\,\Msun$. 
The observed LBGs at $z \lesssim 5$ show metallicities ranging from $\sim 0.1$ to $\sim 
0.4\,\rm Z_{\odot}$ \citep[e.g.,][]{Pettini01, Shapley03, Maiolino08, Nakajima13}. 
Our CR results exhibit similar median metallicities, with only a few outliers having near-solar abundances. 

However, the halos of the massive galaxies in CR have already reached the mass range of $\sim 10^{12}\,\Msun$
by $z\sim 6$, which is similar to that of the observed LBGs at $z < 5$  \citep[e.g.,][]{Ouchi04b}. 
Galaxies in overdense region evolve ahead of the normal regions. Therefore, these massive galaxies can be
substantially enriched by metals during this rapid galaxy growth. 
In addition, there is little information about the metallicity of galaxies at $z > 6$ due to difficulties in 
detecting the metal lines. \citet{Bouwens10} have noted a very low metallicity of LBGs at $z\sim 7$ from the 
extreme blue colors of their UV spectra, while in the recently observed large sample, 
the UV slope of LBGs at $z > 6$ appears similar to that of galaxies at $z\lesssim 5$ \citep{Dunlop12, Bouwens14a}.
Hence, the metallicity of galaxies at $z\gtorder 6$ is still subject to debate. 

The bottom two panels of Figure~\ref{fig:sfr} show that the galactic dust mass, $M_{\rm dust}$, is tightly 
correlated with the stellar mass, because $\sim 0.14$ of the stellar mass ends up
as type-II SNe, and the amount of dust is fixed \citep{Todini01}. On the other hand, some fraction of dust is 
consumed by the SF process or is blown out of galaxies by the SNe feedback. Again, this increases dispersion
in  $M_{\rm dust}$.  At $z\sim 6.3$, the dust mass in the most massive (CR) galaxy has reached $\Mdust \sim 
4.1\times 10^8\,\Msun$. The sSFR and the dust mass in our massive (CR) galaxies are similar to those observed 
in SMGs at $z\sim 3$ \citep[e.g.,][]{Hatsukade11, Michalowski12}. They appear only slightly smaller than in the 
bright SMG at $z = 6.3$ discovered by \citet{Riechers13}, 
which has $\SFR\sim 2900\,\Msun\,{\rm yr^{-1}}$, $\Mst\sim 3.7\times 10^{10}\,\Msun$, and 
$\Mdust\sim 1.3\times 10^{9}\,\Msun$.
The most massive galaxy in our simulations has similar stellar and dust masses, but its SFR is smaller by 
a factor less than 4.  Given the uncertainties in modeling the SF process, this is a very reasonable 
correspondence.

\subsection{UV properties of galaxies}

\begin{figure}
\begin{center}
\includegraphics[scale=0.5]{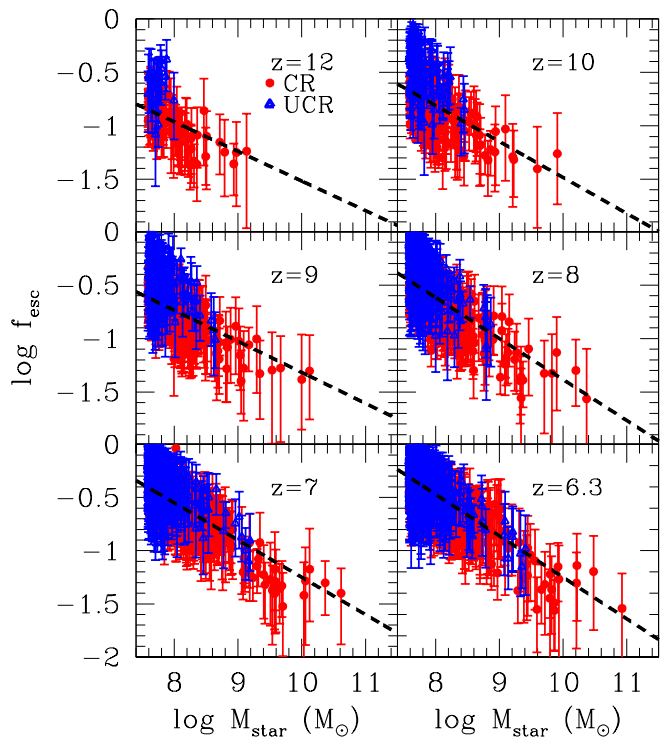}
\caption{
The calculated photon escape fraction, $\fesc$, at $1300\le\lambda\le 2000~{\rm \AA}$ in the rest-frame 
of the galaxy. $\fesc$ is obtained by exact RT calculations and each point represent a 
galaxy in the CR and UCR samples (see text). The dashed lines 
represent the fits to the median values at each mass-bin with the bin-size of 0.5 dex, by using an equation shown 
in Table~\ref{table:fit}. 
The error-bars provide 1-$\sigma$ standard deviations for different
viewing angles. Here, 50 angular bins have been used, 5 in $\theta$ and 10 in $\phi$.
}
\label{fig:fesc}
\end{center}
\end{figure}

\begin{figure}
\begin{center}
\includegraphics[scale=0.63]{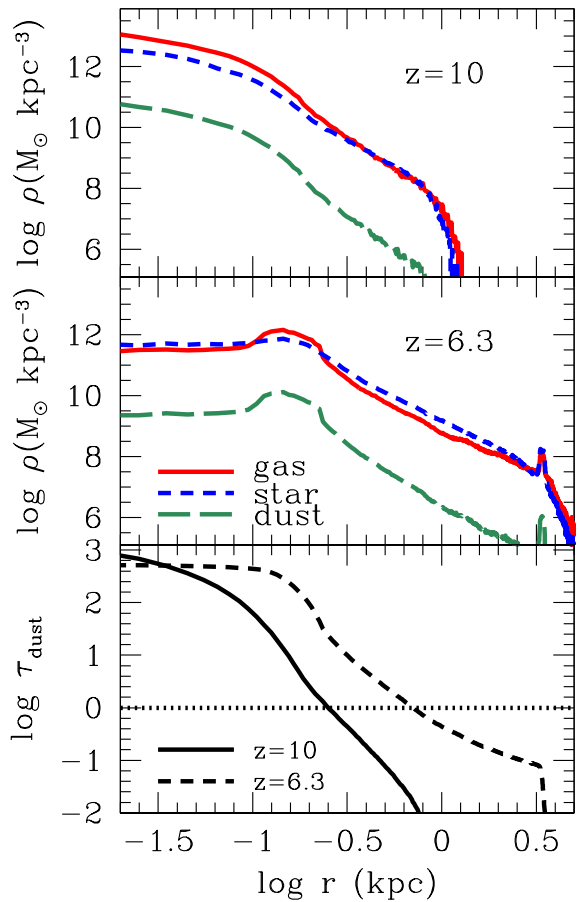}
\caption{
{\it Upper/middle:} The mean densities of gas, stars, and dust as a function of the radial distance
from the galactic centers in physical coordinates. 
The samples consist of the most massive galaxies in the CR run at $z = 6.3$ and $10$.
{\it Lower:} Optical depth of the dust
calculated from the DM halo radius identified by the HOP group-finder algorithm, to a specific 
radius, using density distributions of the upper and middle panels. 
}
\label{fig:dist}
\end{center}
\end{figure}

\begin{figure}
\begin{center}
\includegraphics[scale=0.45]{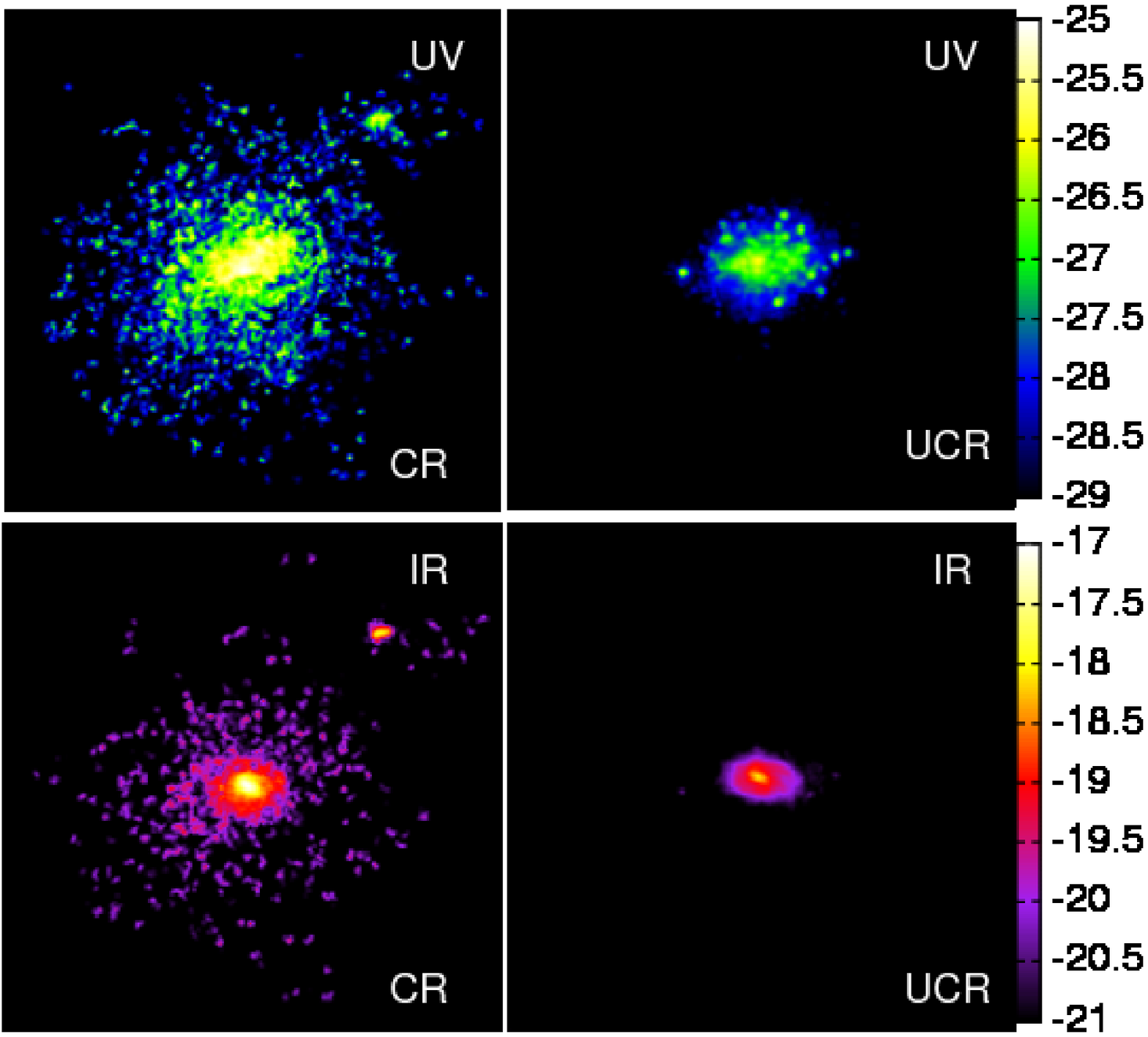}
\caption{
The surface brightness of the most massive galaxies in the CR (left column) and UCR runs (right column) at $z = 6.3$. 
Each panel size is the same as in the Figure~\ref{fig:2dgas}. The upper panels show the UV band ($\lambda\sim 
1600\,\rm \AA$ in the rest frame), which corresponds to the $J$ band in the observed frame.
The lower panels display the IR band ($\lambda\sim 106\,{\rm \mu m}$), 
which corresponds to the sub-millimetre band of $\lambda\sim 850\,{\rm \mu m}$.
The colours indicate the surface brightness in the log scale in units of ${\rm
erg\,s^{-1}\,cm^{-2}\,Hz^{-1}\,arcsec^{-2}}$.
}
\label{fig:sb}
\end{center}
\end{figure}

The UV continuum photons are overwhelmingly emitted by young stars (in the absence of the AGN). Some of 
them are absorbed by the dust before escaping from galaxies. Therefore, the galaxy UV properties are 
determined by their physical properties shown in Figure~\ref{fig:sfr}. 

The calculated escape fraction, $\fesc$, of the UV photons in the $\lambda = 1300 - 2000\,{\rm \AA}$ band
for the CR and UCR galaxies is displayed in the Figure~\ref{fig:fesc}.
We find that most of the UV photons in galaxies with $M_{\rm star}\gtorder 10^8\,M_\odot$ have been absorbed 
by dust already at high redshifts, $z\sim 12$. Their photon escape fraction is $\fesc\ltorder 0.1$ at
this redshift and 
decreases further with the stellar  mass. This mass dependence is consistent with the previous works 
\citep[e.g.,][]{Yajima11, Yajima14c}, 
and the values at $z = 7$ are similar to the published ones \citep[e.g.,][]{Kimm13}.

Figure~\ref{fig:fesc} shows that $\fesc$ is increasing towards smaller redshifts for  
galaxies with $\Mstar\ltorder 10^9\,\Msun$, while for higher-mass objects this increase is marginal.
As a result, the median slope steepens with time. 
Dashed lines fit the median values with the mass bin of $0.5$ dex using the fitting function,
${\rm log}\fesc = \alpha {\rm log}\Mst + \beta$, where $\alpha$ and $\beta$ are the fitting parameters.
The slope $\alpha$ ranges from $\sim -0.28$ to $\sim -0.39$.

The benchmark value of $\fesc\sim 0.1$ moves to the right and is located at 
$M_{\rm star}\sim 3\times 10^9\,M_\odot$ by $z\sim 6$. 
The parameter $\beta$ evolves from $1.25$ at $z = 12$ to $2.65$ at $z=6.3$.
For galaxies with $M_{\rm star}\sim 10^{10}\,M_\odot$, the $\fesc$ 
increases from 0.03 to 0.06 during $z\sim 10-6.3$, while for $M_{\rm star}\sim 10^8\,M_\odot$ galaxies
it rises by a factor of 3. Hence galaxies become more transparent at lower $z$ in our simulations,
and low-mass galaxies show a more pronounced trend in this direction.

The $\fesc$ in our simulations is smaller than that of \citet{Dayal13}. 
These authors have assumed that dust is distributed in a screen with a radius of $0.225R_{200}$
where $R_{200}$ is the halo virial radius, taking the mean density of the sphere to be 200 times the critical 
density at a particular redshift. 
On the other hand, in our simulations, the gas distribution is clumpy, is found near the starforming regions, 
and, therefore, absorbs the stellar radiation more efficiently, resulting in lower $\fesc$.

The observed LBGs indicate a very high escape fraction because of the blue colours in the UV bands
\citep[e.g.,][]{Bouwens10, Dunlop13}.  
In comparison, the $\fesc$ of high-mass end galaxies in our 
simulations appears smaller. By $z\sim 6.3$, our most massive CR galaxies with $\sim 10^{11}\,M_\odot$
have $\fesc\sim 0.03$.
Note that the estimated higher $\fesc$ in the observed LBGs result from the assumed smooth dust extinction curve
of local starforming galaxies from \citet{Calzetti00}. However, if one accounts for the bump at $\lambda = 
2175~\rm \AA$ in the dust extinction curve, the blue colour could still be obtained, despite $\fesc$ being small.
 
Massive CR galaxies are compact, and have large amount of dust due to the high 
SFR. The stellar distributions are centrally concentrated, and the galactic centers --- the sites of the 
SF --- appear to be enshrouded in dust. Hence, the stellar radiation is efficiently absorbed. We note, that
there is uncertainty in $\fesc$ when using the UV colour alone, 
because the change in the UV colours by the dust extinction depends sensitively on the dust properties, 
e.g., its size and composition.  If the typical grain size is smaller than that in the Milky Way, and the main 
component is graphite, the dust extinction does not change the UV colour or make it bluer because of the 
bump at $\lambda = 2175$\,\AA\ in the dust extinction curve \citep{Inoue06, Kimm13}.

The most significant difference in $\fesc$ between the CR and UCR galaxies, less than a factor of 2, can be 
seen in the lowest mass bin, $M_{\rm star}\sim 10^{7.5}-10^{8.5}\,M_\odot$, as we discuss in Section 3.5. The 
reason for this is that the gas distribution, and hence the distribution of SF regions and 
dust, in the CR galaxies are more compact  than in the UCR ones. 
As shown in \citet{Romano-Diaz14}, the galaxies in a particular mass range form earlier in overdense regions,
compared to `normal' regions in the universe. This happens because the accretion rate depends on the background
density in addition to the halo/galaxy mass.
Indeed, in our simulations, the low-mass galaxies in CR form earlier than the UCR ones, resulting in a more
compact gas/star distribution and a more efficient dust extinction.
At earlier times, the universe is denser and the parent DM halos have larger concentration parameter, causing 
compact gas distribution.

The radial distribution of the gas, stars and dust in the CR run are shown in Figure~\ref{fig:dist}.
We take the most massive galaxies at $z=6.3$ and $10$, 
and estimate the mean density in spherical shells at each radius. 
Most of them are distributed in the central kiloparsec. Half-mass radii of the gas are 95\,pc 
at $z=10$ and 215\,pc at $z=6.3$. Due to the compact distributions, the optical depth at $\lambda=1300\,\A$, 
estimated from the boundary of the grid placed around each galaxy 
to an arbitrary radius $r$, i.e., $\tau(r) = \int_{\rm boundary}^{r} 
\sigma_{\rm 1300} \rho_{\rm dust}(r)dr$, increases steeply when approaching the galactic centre.  
At $z=10$, the optical depth to $r\ltorder 0.25$\,kpc is larger than unity, and this radius increases to 
$r=0.71$\,kpc at $z=6.3$. The stellar distribution behaves similarly.  For example, the mass $\sim 
0.8\,M_{\rm star}$ is 
within the radius mentioned above, where the ISM is optically thick at $z=10$, but decreases somewhat
to $\sim 0.68M_{\rm star}$ at $z=6.3$. 

Consequently, most of the stellar UV radiation is absorbed by the dust. 
Note, that, despite lower dust mass at $z=10$, the fraction of stellar mass inside the radius of $\tau=1$ is 
higher than that at $z=6.3$, due to the compact dust and stellar distributions. Therefore, even at high 
redshifts, stellar UV radiation of galaxies residing in the overdense regions can be absorbed by the dust. 
The gas and dust appear to be pushed away from the center at lower redshifts.

The surface brightness of the most massive galaxies in the CR and UCR runs at $z=6.3$ is shown in $J$ and 
sub-millimetre bands (Figure~\ref{fig:sb}). 
The $J$ band in the observed frame corresponds to the UV band in the rest frame.
The $J$-band flux of the galaxy in the CR run exhibits a similar 
distribution to the stellar one (Figure~\ref{fig:dist}), e.g., 1-D radial distribution. However, 
some parts get fainter, because of the stronger dust absorption.
On the other hand, the sub-millimetre 
flux traces the dust distribution, and, therefore, is more compact compared to the $J$-band flux. 
The $J$ and sub-millimetre band fluxes of the galaxy in the UCR run show similar distributions reflecting 
stars and dust, respectively.
However, the UV flux is flatter compared to the stellar distribution, and the sub-millimetre flux is more 
compact due to the effect of the radiative transfer, i.e., 
UV flux near the galactic centers is efficiently absorbed, as shown in Figure~\ref{fig:dist}. 

\begin{figure*}
\begin{center}	
\includegraphics[scale=0.8]{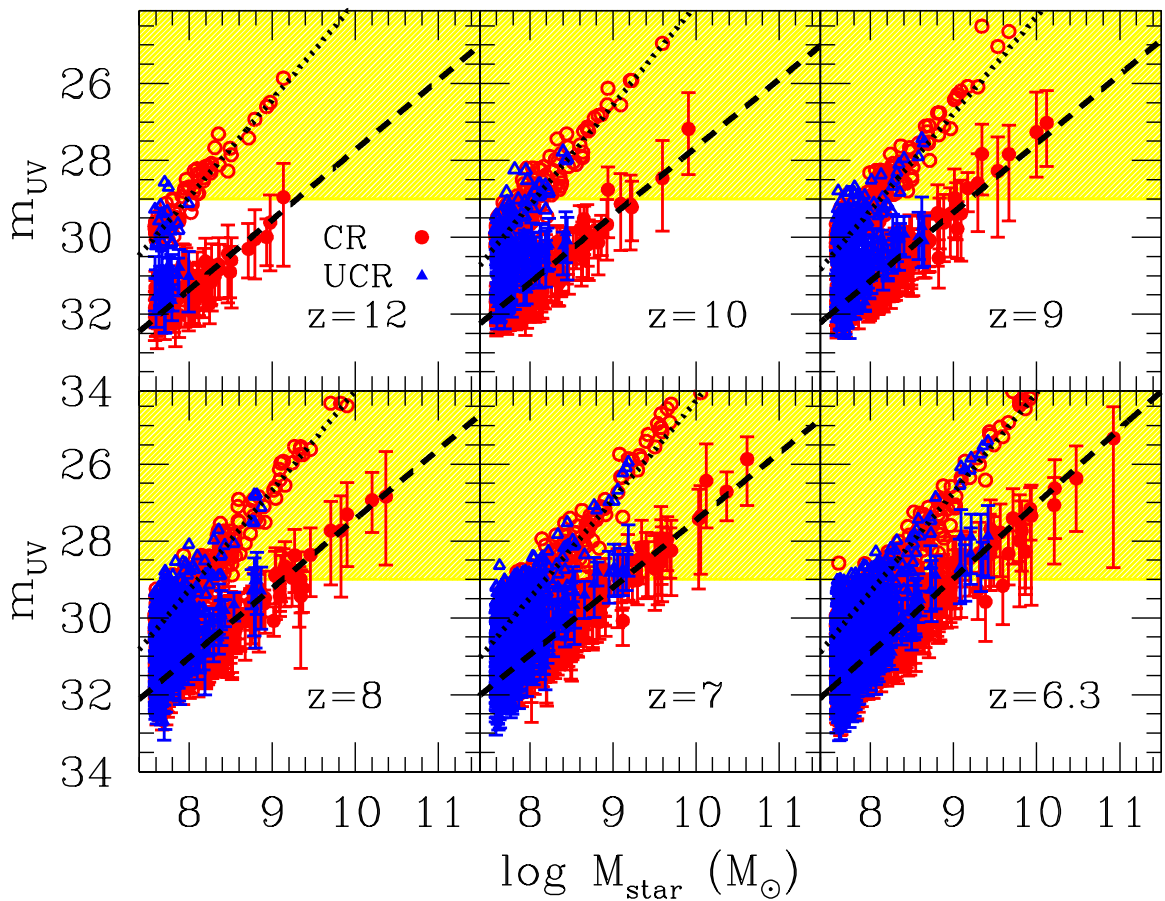}
\caption{
The UV flux density in AB magnitudes as a function of stellar mass, $M_{\rm star}$. 
The yellow shades show the detectable region ($m_{\rm UV}\sim 29\,\rm mag$) in the recent observations using 
the HST \citep[e.g.,][]{Bouwens11b}. The red and blue symbols show galaxies in the CR and UCR runs 
respectively. Filled and open symbols represent the UV fluxes with and without the dust extinction. 
The error bars of the filled circles represent the minimum and maximum values of the different viewing 
angles of a galaxy. The dotted and dashed lines are fitting lines to the median values with a bin size of 
0.5\,dex with and without the dust extinction, when applying the fitting function $m_{\rm UV} = \alpha\, 
{\rm log}\,M_{\rm star} + \beta$, where $\alpha$ and $\beta$ are the fitting parameters (see text).
}
\label{fig:muv}
\end{center}
\end{figure*}

The UV fluxes from galaxies at $\lambda\sim 1300\,\A$ in the rest frame are shown in Figure~\ref{fig:muv}. 
The flux is the angular mean. Dotted and dashed lines represent fitting lines to median values with the mass 
bin of 0.5\,dex, using the 
function $\muv = \alpha\,{\rm log}\,\Mst + \beta$. For the intrinsic $\muv$, $\alpha\sim -2.3$,
which means that the UV flux increases linearly with the stellar mass, $M_{\rm star}$, since $\muv\propto 
-2.5\,{\rm log}\,F_{\rm \nu}^{\rm UV}$, where $F_{\rm \nu}^{\rm UV}$ is the UV flux density. 
This happens because the UV flux is proportional to the SFR \citep{Kennicutt98} which correlates with the 
stellar mass, as shown in Figure~\ref{fig:sfr}. On the other hand, due to the mass dependence of $\fesc$, 
the fitting lines to $\muv$ are flatter with $\alpha \sim -1.3$,  considering the dust extinction. 
Hence, after correction by the dust extinction, $F_{\rm \nu}^{\rm UV} \propto M_{\rm star}^{0.52}$.

We find that, before the dust extinction has been applied, many CR galaxies appear above the detection 
threshold for recent observations of the Hubble Ultra-Deep Field (HUDF) with the Wide-Field Camera\,3 (WFC3) 
onboard the HST   \citep{Bouwens11b}. 
However, due to the strong dust extinction, the galaxies in the CR run become fainter by $\sim 5$ magnitudes. 
As a result, many of the galaxies appear below the detection threshold for the current HST observations. 
At $z\sim 10$, only a few bright galaxies appear above the threshold. 
In the UCR run, the intrinsic UV fluxes of most galaxies are fainter than the 
detection threshold at $z \gtrsim 8$. Only a few galaxies at $z\lesssim 7$ can be observable as LBGs.

\begin{figure}
\begin{center}
\includegraphics[scale=0.4]{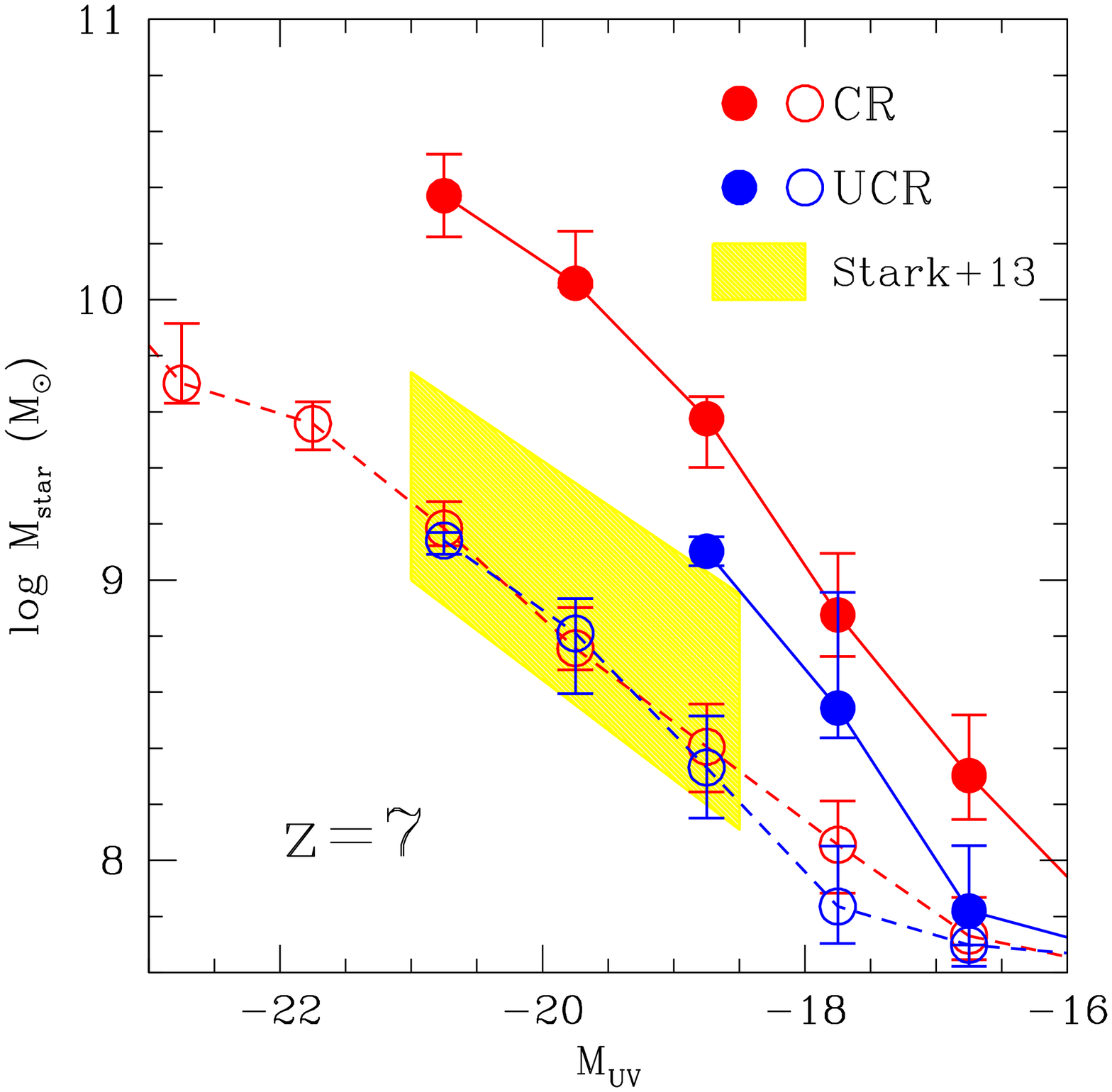}
\caption{
Stellar masses of galaxies as a function of the rest-frame UV magnitude at z=7. 
Filled and open circles represent median values of galaxies with corrected and uncorrected UV fluxes 
by dust extinction, respectively. The yellow shaded region shows the observed galactic stellar masses 
accounting for the ambiguity from the contribution by the nebular lines from \citet{Stark13}.
}
\label{fig:mstz7}
\end{center}
\end{figure}

The error bars in Figure~\ref{fig:muv} show the dispersion in $\muv$ due to the viewing angle. Because of 
the anisotropic stellar and dust distributions, the UV flux changes significantly. 
On average, the dispersion increases with the stellar mass, because 
the galaxy shapes become more disky \citep[see \S\,~\ref{sec:view}, and also][]{Romano-Diaz11b, Biffi13}, 
and the photons escape efficiently along the 
normal direction to the disk plane. This can affect significantly the limiting CR galaxy mass, moving it
by about a decade in $M_{\rm star}$, from $\sim 10^9\,M_\odot$ to $\sim 10^{10}\,M_\odot$.
For example, $\muv$ of the most massive galaxy at $z=6.3$ in the CR 
ranges from $24.5$ to $28.7$. Therefore, for many galaxies, the detection by the HST depends on the viewing 
angle. Since the dust absorption cross section decreases with increasing wavelength, dispersion 
becomes smaller at longer wavelengths. This effect may cause variation of physical properties in 
observational study using the SED fitting. We shall discuss additional details in 
Section~\ref{sec:discussion}.

Recently, \citet{Stark13} estimated the relation between stellar mass and $M_{\rm UV}$ at $z \sim 4-7$. 
Figure~\ref{fig:mstz7} shows the relationship between the stellar masses of galaxies and their emerging UV 
flux in our simulations at $z=7$, corrected and uncorrected for the dust attenuation. 
Stellar masses in the UCR galaxies lie within the range estimated by these observations. 
Because the UCR run is using a small volume, we cannot sample larger galaxies, because they are rare and 
difficult to find in such volumes (see 
Section\,2). The stellar masses of our CR galaxies with $M_{\rm UV} > -18$ exceed those of Stark et al. by a 
factor of a few, while the dimmer ones fall within the observed range.
We note that the CR galaxies reside in the highly overdense region which tends to evolve ahead of the
average density regions in the universe. For example, their SF starts at higher redshifts than in the UCR
galaxies. As a result, a larger fraction of the UV photons will be attenuated by larger amount of dust, and 
the stellar mass-to-emerging UV flux 
ratio becomes higher in comparison with the observed LBGs. 
Our numerical model implementation has also successfully reproduced the cosmic SFR density, the metal abundance 
in the IGM \citep[e.g.,][]{Choi09}, 
and the neutral hydrogen abundance \citep[e.g.,][]{Nagamine10, Yajima12d}.
Furthermore, observations of stellar masses in galaxies include uncertain parameters, such as ages, 
star formation history, dust extinction, and nebular emission lines in the SED fitting. 
These parameters could change the estimated values of stellar masses significantly, and make it more in 
agreement with our CR models. Given these considerations, we conclude that the stellar masses of our simulated 
galaxies are in reasonable agreement with those observed by \citet{Stark13}.

\subsection{UV luminosity function of galaxies}
\label{sec:LF}

 
\begin{figure*}
\begin{center}
\includegraphics[scale=0.8]{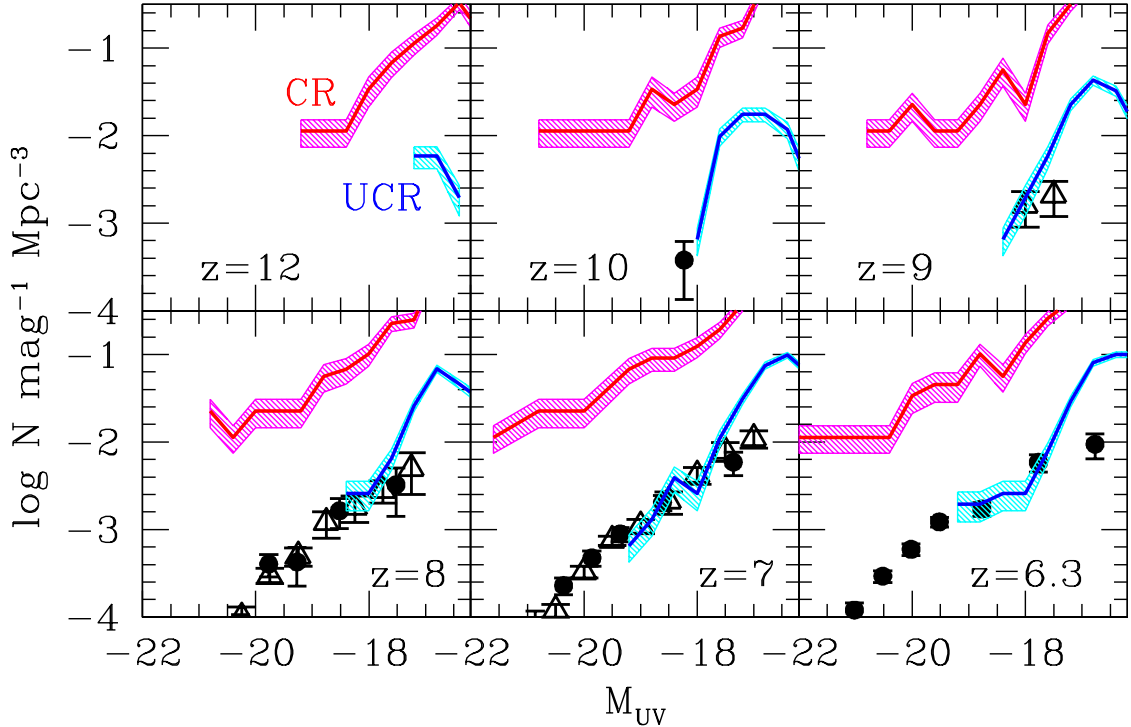}
\caption{
The galaxy UV luminosity function after applying the dust correction (see text). 
The red and blue lines represent the simulation results for the CR and UCR runs, respectively.
The shaded regions represent the Poisson errors.
Filled circles and open triangles are observational data by \citet{Bouwens14} at $z=6, 7, 8,$ and 10, 
and \citet{McLure13} at $z=7, 8,$ and 9, respectively.   
}
\label{fig:LF}
\end{center}
\end{figure*}

\begin{figure}
\begin{center}
\includegraphics[scale=0.5]{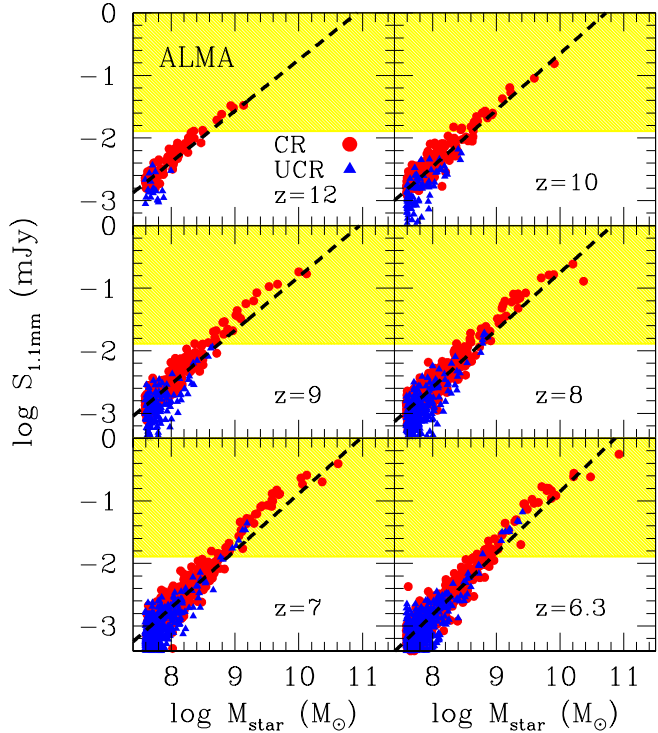}
\caption{
The sub-millimetre flux at $\lambda = 1.1\,{\rm mm}$ in the observed frame. 
The yellow shades show the region of $S_{\rm 1.1mm} > 0.013 \,{\rm mJy}$, which corresponds to 
the detection threshold of ALMA with $\sim 1$\,hour integration using 50 12-m antennas. 
}
\label{fig:submm}
\end{center}
\end{figure}

Galaxy luminosity functions (LFs) range has been extended recently to $z\gtorder 8$ from observations in
the UDF, GOODS-N, GOODS-S and the XDF fields \citep{Bouwens14}. Moreover, using the UKIDSS field, 
few candidates have been detected also at $z\gtorder 9$ \citep[][see also Oesch et al. 
2013]{McLure13}.
This enables us to compare our simulation results to the observed LFs as shown in Figure~\ref{fig:LF}.

Since the CR simulation has been performed in a targeted, highly overdense environment of a $5\sigma$  
region, the number densities of galaxies are substantially higher than the mean ones probed by 
the observations. At present, the Constrained Realization method constitutes an elegant way to
sample such statistically rare regions without any loss of generality, compared to the use of large-scale
simulations with re-sampling.   

On the other hand, we also demonstrate that the LF in the UCR run nicely agrees with the observed LFs at the 
fainter side. 
Without the dust extinction (not shown here), the LFs are shifted a few magnitudes brighter, and hence 
deviate from the observed ones. These shifts are related to $\fesc$ through
$M_{\rm UV} - M_{\rm UV}^{\rm int} = -2.5\; {\rm log}\,\fesc$. 
Therefore, $M_{\rm UV}$ of massive CR galaxies with $\fesc < 0.1$ becomes fainter by more than 2.5 
magnitudes, while that of the low-mass UCR galaxies with $\fesc > 0.3$ shifts down by less than
1.3 magnitudes. 

This result is interesting and of a prime importance to observations of high-redshift 
galaxies, as it shows that the dust extinction is important in the $z\gtorder 6$ galaxies.
We note that due to a relatively small computational box for the UCR galaxies, it does not include massive bright 
galaxies with $L > L^{*}$. Hence, our results in the UCR run are limited at $m_{\rm UV} \gtrsim -19$.

The reasonable agreement of the UCR LF with observations after the dust extinction correction also
provides support to the CR run results, which has been performed with the same physical models, except
for the environment of higher densities. We note that, if quasars indeed reside in such massive galaxies, 
their spatial distribution will reflect the distribution of our CR galaxies in the sky with the limited volume, 
accounting for quasar duty cycle somewhat smaller than unity.

Our results indicate that even galaxies at $z\gtorder 6$ can suffer from the dust attenuation. 
On the other hand, recent observations have shown that the slope $\beta$ of the UV flux is 
$\sim -2$ \citep[e.g.,][]{Dunlop12}. 
If we consider the smooth dust extinction curve, such as the Calzetti law \citep{Calzetti00}, 
$\beta \sim -2$ means that the dust attenuation is small \citep{Meurer99}. 
However, the SNe dust model is associated with the extinction curve which
exhibits a bump at 2175\,\AA, due to the small graphite dust. 
Therefore, our SNe dust model does not increase $\beta$ with the dust extinction,
i.e., the dust extinction $A_{\rm UV}$ is nearly independent of this parameter 
\citep[e.g., Figure~1 in][]{Yajima14b}. 
This is unlike the relation between the dust extinction and the slope $\beta$ in the local starforming galaxies 
\citep{Meurer99}.  We confirm that even galaxies with a strong dust extinction keep $\beta\sim -2$, 
which is nearly the value of the intrinsic SEDs.
We also note that the UV slope through broad-band filters can change significantly
by the contribution from  nebular lines \citep[e.g.,][]{Schaerer03, Schaerer09},
which is not included in our current simulations.
 
\subsection{Infrared properties of galaxies and detection by ALMA}
\label{sec:IR}

Dust absorption of the stellar radiation leads to the dust thermal emission in the infrared band, 
in the galaxy's rest frame. 
This emission from the high-redshift galaxies is bright in the sub-millimetre band of the observed 
frame. Figure~\ref{fig:submm} displays the sub-millimetre fluxes at 1.1 mm, hereafter $S_{\rm 1.1}$. 
The SFR and the dust mass increase linearly with the galactic stellar mass as seen in Figure~\ref{fig:sfr}. 
Most of the stellar radiation is absorbed, therefore, it leads to the increasing sub-millimetre and the IR 
luminosity, $L_{\rm IR}$, with increasing $M_{\rm star}$. The most massive galaxies in the CR run have 
$L_{\rm IR}\sim 6.3\times 10^{11}\,L_{\odot}$ at $z\sim 10$ and $3.7\times 10^{12}\,L_{\odot}$ at 
$z = 6.3$. The dash lines are fitted to median values with the mass bin of $\Delta\,{\rm
log}\,M_{\rm star} = 0.5$, using a function ${\rm log}\,S_{\rm 1.1} = \alpha\,{\rm log}\,M_{\rm
star} + \beta$.
The IR luminosity displays a similar slope to that of the SFR, namely $L_{\rm IR}\propto
\Mst^{1.1}$. This happens because the UV luminosity, which is given by the SFR, is nearly
all converted into $L_{\rm IR}$.
On the other hand, the slope of $S_{1.1}$ shows a shallower $\alpha\sim 0.9$,
because  $\lambda = 1.1\,{\rm mm}$ lies at the long-wavelength tail of the thermal peak of the dust 
emission. For example,  
the peak wavelength of IR radiation from the most massive galaxy
is $\sim 300\,{\rm \mu m}$, 
while that of the observed SMG shows $\sim 500\,{\rm \mu m}$.
Therefore, when the dust temperature increases with increasing $L_{\rm IR}$, the peak
intensity moves to smaller $\lambda$ and becomes higher. Hence the increase in the intensity
around the peak is larger than in the tail. 
For a fixed covering area of dust emission, 
this translates into 
slower increase in the IR flux at the tail, compared to $L_{\rm IR}$.

The detection threshold by the recent ALMA observations is $S_{1.1}\sim 0.013 \,{\rm mJy}$, which 
corresponds to $\sim 1$\,hour integration with 50 12-m antennas. 
Even at $z = 8$, the galaxies in the UCR run appear undetectable because of the lower SFR. 
On the other hand, massive galaxies in the CR run can be easily detected by ALMA up to $z\sim 10$. 

The massive galaxies in CR run appear fainter than the observed SMG at $z=6.3$ by factor $\sim 10$ 
\citep{Riechers13}, although the stellar and dust masses are similar. 
There are at least three options which can explain such a difference. 
First, the observed SMG could be detected during an extreme starburst phase, 
while our model galaxies mostly evolved with smooth gas accretion \citep{Romano-Diaz14}.
Second, the CR galaxies could have a different dust covering factor of the central SF sites.
Alternatively, the
modelled and observed galaxies can lie in different environments, which lead to different accretion
rates, gas fractions, etc. Note that the number of detected SMGs at $z\gtrsim 6$ is still quite small. 
Future surveys of high-redshift SMGs using ALMA will allow for a better and statistically significant
comparison. 

While we focus on the dust continuum emission in this work, we note that the metal lines, e.g., C and  O, and 
the CO molecular lines enhance the detectability of high-redshift galaxies \citep[e.g.,][]{Inoue14, 
Tomassetti14}. 

\subsection{Redshift evolution of galaxies}
\label{sec:redshift}

\begin{figure}
\begin{center}
\includegraphics[scale=0.43]{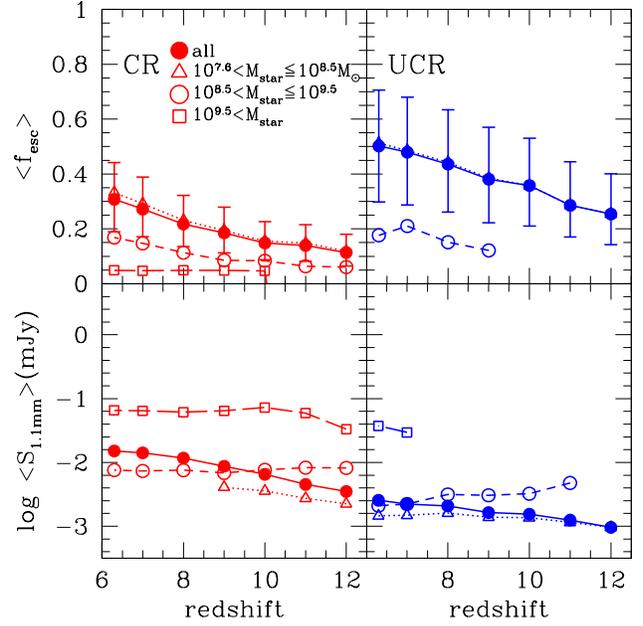}
\caption{
{\it Upper:} Redshift evolution of the UV escape fraction $\fesc$. Filled symbols represent the median 
values of the full sample, open symbols are the means in each mass bin. Red and blue 
colours 
represent the CR and UCR runs, respectively. Note that red and blue triangles follow the ``all'' red
and blue filled circles. The error bars are the quartiles of the sample in each mass bin. 
{\it Lower:} Redshift evolution of the sub-millimetre flux $S_{1.1}$ at $1.1\,{\rm mm}$ in 
the observed frame. 
}
\label{fig:fescz}
\end{center}
\end{figure}

The dust mass increases with time as well as with the SF (Figure~\ref{fig:sfr}), and the dust 
distribution around the star-forming regions changes with time as well 
(Figure~\ref{fig:dist}). Consequently, the spectral properties of galaxies are expected to evolve with 
redshift. Figure~\ref{fig:fescz} shows the redshift evolution of $\fesc$ and the $S_{1.1}$ flux. 
The symbols represent the medians in galaxies in each mass range.
Because of a higher number of low-mass galaxies in the galaxy mass function, the median values are 
similar to those of low-mass 
samples ($10^{7.6} < \Mst \le 10^{8.5}\,\Msun$). This effect dominates the UCR run because the massive 
galaxies are completely absent there. The median values increase with decreasing 
redshift, despite of the increase in the dust amount. 

As shown in Figure~\ref{fig:dist}, at higher redshift, the dust is distributed compactly around the 
starforming region at the galactic centre. 
Therefore, smaller amounts of dust can efficiently absorb the stellar radiation, resulting in the 
suppression of $\fesc$.
As we have discussed in Section\,3.3, $\fesc$ of the CR galaxies in the lowest mass bin lies below 
that of the UCR one, by less than a factor of 2 (see Figure~\ref{fig:fescz}). 
The photon escape fraction,
$\fesc$, of the high-mass galaxies with $\Mst\gtorder 10^{9.5}\,\Msun$, does not change much.
These galaxies have extended disks even at high redshift of $z \gtrsim 10$ \citep{Romano-Diaz11b}, 
hence photons can escape from stars at the outskirt of a galaxy. 

The bottom panels show the redshift evolution of $S_{1.1}$.
Although $L_{\rm IR}$ increases with decreasing redshift because of the higher SFR, 
$S_{1.1}$ does not change significantly. 
Due to an increase in the dust temperature with $L_{\rm IR}$, 
the $S_{1.1}$ flux is nearly constant. 
Only massive galaxies with $\Mst\gtorder 10^{9.5}\,\Msun$ show an increase in $S_{1.1}$ by a factor of
2 between $z\sim 12$ and 10, and can be detected by ALMA, 
while most galaxies are too faint with $S_{1.1} \lesssim 10^{-2}\,{\rm mJy}$.

\subsection{Effect of the galactic morphology on the spectral properties}
\label{sec:view}

Escape of the UV continuum photons from galaxies is sensitive to the viewing angles due to the anisotropic 
column densities, as shown in Figures~\ref{fig:fesc} and \ref{fig:muv}.
Here we investigate the effect of galaxy morphology on the dispersion of $\fesc$.
We define the three galactic semi-axes as $a$, $b$ and $c$ from the inertia tensor, 
with $a > b > c$ \citep{Heller07}.

\begin{figure}
\begin{center}
\includegraphics[scale=0.46]{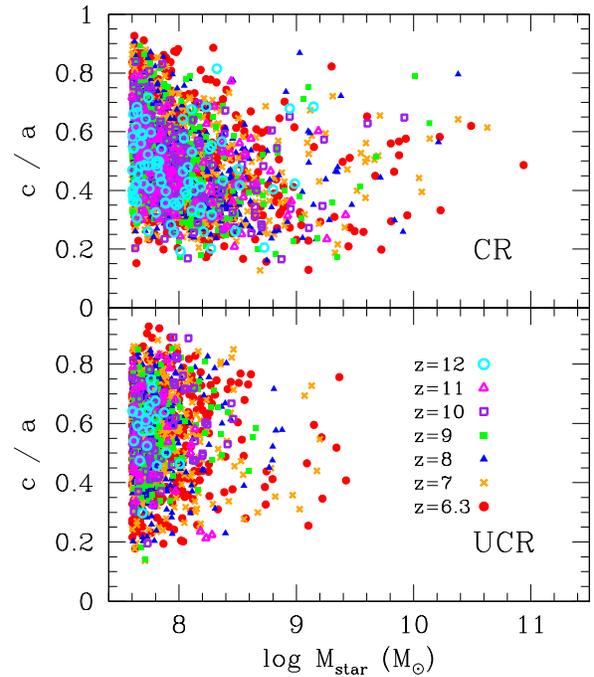}
\caption{
The axial ratio $(c/a)$ as a 
function of stellar mass for the CR 
and UCR runs. See definitions of the galactic semi-axes, $a$, $b$, and $c$ in the text.
}
\label{fig:mstshape}
\end{center}
\end{figure}

The relation between $c/a$ and stellar mass is shown in Figure~\ref{fig:mstshape}. 
No clear correlation can be seen for $\Mst \lesssim 10^{8}~\Msun$ for the CR and UCR runs.
Only the CR run shows that $c/a$ of massive galaxies decreases roughly with increasing stellar masses.
This can be directly related to the high accretion rate of cold gas from cosmological filaments around 
high-density peaks, which tends to form an extended disk \citep[see also][]{Romano-Diaz14}. 
These disks appear to be resilient (Romano-Diaz et al., in preparation). 
In addition, the dispersion becomes large for $\Mstar\gtorder 10^9\,\Msun$ again, 
this might be due to merging processes. 

\begin{figure}
\begin{center}
\includegraphics[scale=0.45]{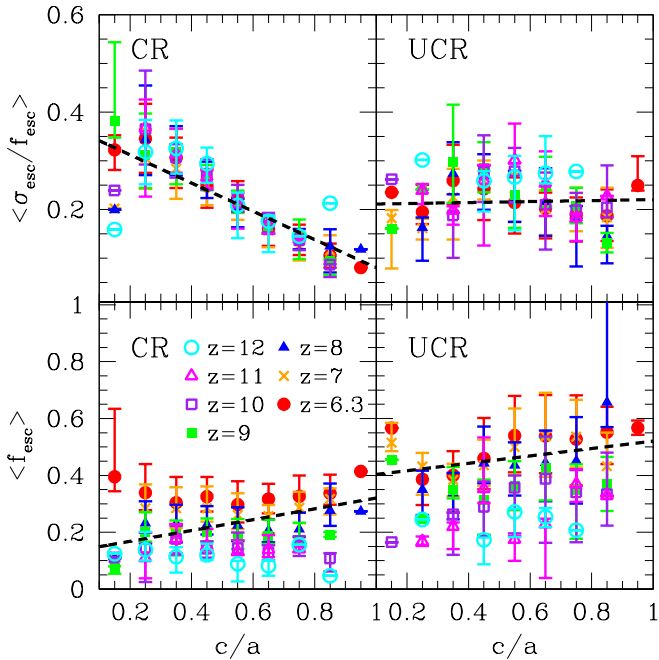}
\caption{
{\it Upper:} 
Coefficients of variation of the UV escape fraction for different viewing angles normalized 
by the angular mean values, $\equiv\sigma\,/\,\fesc$, as a function of the galactic axial ratio $c/a$.
Symbols represent medians at each bin.  The error bars are top and bottom quartiles. 
{\it Lower:} Escape fraction of the UV continuum photons as a function of $c/a$.
The dash lines represent fit to the median values for all redshift samples by using a function
$\sigma_{\rm esc}/\fesc$, or $\fesc = \alpha\,(c/a) + \beta$,
where $\alpha$ and $\beta$ are the fitting parameters. 
}
\label{fig:fescshape}
\end{center}
\end{figure}

Then we measure the $1\sigma$ error in $\fesc$ by different viewing angles, $\sigma_{\rm esc}$, normalised by the 
angular mean $\fesc$ of each galaxy, as a function of $c/a$.
Figure~\ref{fig:fescshape} displays the median values of each redshift sample.
Since most galaxies with $M_{\rm star}\gtorder 10^9\,\Msun$ show $c/a << 1$ (and $a\sim b$), this leads 
to a verifiable
conclusion about the presence of geometrically-thin disks in these objects. Spheroidal galaxies show 
$c/a\sim 1$ and $a\sim b$. For the CR run, scatter in $\fesc$ increases roughly with decreasing $c/a$
(e.g., upper left frame in Figure~\ref{fig:fescshape}). 
Photons can escape along the normal direction to the disk plane due to the lower column density.

The ratio $\sigma_{\rm esc}/\fesc$ ranges from  $\sim 0.35$ at $c/a\sim 0.2$ to 
$\sim 0.1$ at $c/a\sim 0.8$.
The median values can be fit by $\sigma/\fesc = -0.29\times (c/a) + 0.37$.
On the other hand, for the UCR run, most galaxies are low-mass, and the optical depth along the edge-on 
disks is not so high. Therefore, even disk galaxies do not show large dispersion for the UCR run.
Furthermore, some galaxies with $c/a\sim 0.5$ have high $\sigma_{\rm esc}/\fesc$ and $\fesc$.
We measure the Pearson's correlation coefficient 
for all samples shown in the top panels of Figure~\ref{fig:fescshape}, 
and obtain $-0.66$ for the CR run and $-0.18$ for the UCR run. 
This clearly shows that the correlation is stronger for the CR galaxies in the top-left panel. 

The bottom panels of Figure~\ref{fig:fescshape} show $\fesc$ as a function of $c/a$.
Here, $\fesc$ is the angle-averaged value, and each point represents one galaxy.
There appears to be no clear correlation between the morphology and $\fesc$.
The correlation coefficient is $0.03$ for the CR run, and $0.20$ for the UCR run, respectively. 
Thus, we conclude that the disky shape of massive galaxies causes the large dispersion in $\fesc$
value, but does not change $\fesc$ itself significantly.   
In our model galaxies, the high-density gas clumps are distributed around the star-forming regions, 
and efficiently absorb UV photons. Hence, $\fesc$ depends on the covering factor of the clumps, 
and $\sigma_{\rm esc}/\fesc$ is affected by the anisotropy in this distribution. 
Figure~\ref{fig:fescshape} shows that the anisotropy increases with decreasing $c/a$,
but the covering factor does not depend on it significantly. 
For the disk and no-disk galaxies, photons emitted by stars far from the galaxy centers, escape efficiently. 
Yet, there is a preferred direction of escape for photons emitted close to the centers of disk galaxies, i.e., 
the normal direction to the disk plane, resulting in larger dispersion of $\fesc$.

%
%

\section{Discussion \& Summary}
\label{sec:discussion}

We used the high-resolution zoom-in cosmological simulations of galaxies by 
\citet{Romano-Diaz11b,Romano-Diaz14} and have post-processed them with a 3-D radiation transfer code. 
We examined the evolution of multi-wavelength spectral properties of high-redshift galaxies. 
The spatial resolution in the comoving coordinates is $\epsilon = 300\, h^{-1}$pc, 
which corresponds to 
(in physical coordinates) 23\,$h^{-1}$pc at $z\sim 12$ and 43\,$h^{-1}$pc at $z = 6$. 
We used two different types of simulation setup.  
One simulated an unconstrained average region in the universe (the UCR run). 
Another simulation was prepared by imposing a constraint in order to simulate
a rare, highly-overdense, $5\sigma$ region in a cubic box of $20\,h^{-1}$\,Mpc, which includes the 
seed of $\sim 10^{12}\,h^{-1}\Msun$ DM halo projected to collapse by $z\sim 6$ \citep[see][for more details] 
{Romano-Diaz11b, Romano-Diaz14}.  
This was prepared using the Constrained Realization (hereafter the CR run) method \citep{Hoffman91, Romano-Diaz11a}.
Our simulations contain $\sim 50-600$ galaxies $\Mst\gtorder 4\times 10^7\,\Msun$ for each of the seven 
representative snapshots during $z\simeq 6.3-12$. 
The total samples consist of 2273 galaxies for the CR run, and 1789 galaxies for the UCR run. 
 
We find that, in the CR run, massive galaxies with $\Mst\sim 10^{11}\,\Msun$ form 
within DM halos of $\sim 1-2\times 10^{12}\,M_\odot$ by 
$z\sim 6$, and such regions can serve as a hotbed for the formation of high-$z$ quasar hosts --- the 
overall evolution in the overdense region has been accelerated dramatically with respect to the normal 
fields. Numerical simulations estimate that a comoving density of such massive DM halos
exceeds the comoving density of high-$z$ quasars. 
Using the WMAP9 cosmological parameters \citep{Hinshaw13}, we obtain about 200
such halos per $1 h^3\,{\rm Gpc^{-3}}$. 
This makes their comoving space density about 100 times above that of $z\gtorder 6$ quasars, $\sim 2h^3\,{\rm 
Gpc}^{-3}$ \citep[e.g.,][]{Fan.etal:03,Willott.etal:10}, if the duty cycle of these quasars is unity. For
lower duty cycle, their space densities can be similar.
The physical properties of these galaxies change
significantly with their environment, leading to a different dust extinction, and therefore, to different 
spectral appearances.
Much lower mass galaxies with $M_{\rm star}\lesssim 3\times 10^9\,\Msun$  
have been found in the average density run, but this lower limit of the galaxy mass is of course expected to 
increase for larger computational boxes with the same SPH mass resolution.
With our goal of separating the environments of overdense and average regions in the universe,
the adopted sizes of computational boxes are suitable to probe massive and low-mass starforming galaxies for 
the current numerical resolution.

We find that galaxies in the CR and UCR simulations exhibit the following physical properties:

\begin{itemize}
\item The LFs for $M_{\rm UV}\gtorder -19$ of the UCR galaxies at $z \sim 6 - 10$, processed with RT and dust 
extinction, have matched the observed LFs of \citet{Bouwens14} and \citet{McLure13}.  
This is an important test for the modelled galaxies. 
Other than the assumption that 
the dust size follows the SNe dust model and 
the dust distribution traces the metals and implying
that the dust-to-metal mass ratio in these galaxies do not differ from those at low redshifts (but see
Gallerani et al. 2010), no additional
assumptions and no additional corrections have been made. 
It also implies that the dust extinction is important in high-$z$ galaxies, at least to $z\sim 9$, for our 
simulations. 
The LFs for the CR galaxies have been calculated for  $M_{\rm UV}\gtorder -19.5$ at $z=12$,
$M_{\rm UV}\gtorder -21$  at $z = 8-10$, and  $M_{\rm UV}\gtorder -22$  at $z = 6.3$,
and overestimated the observations because of the biased high-density region.

\item The escape fraction of the non-ionizing UV photons, $\fesc$, decreases with increasing stellar mass.
Massive galaxies with  $\Mst\gtrsim 10^{9.5}\,\Msun$ exhibit  $\fesc\ltorder 0.1$, and show little redshift
evolution. While the low-mass galaxies with $\Mst\ltorder 10^{8.5}\,\Msun$ show a strong redshift evolution,
and increase $\fesc$ from $\sim 0.1$ at $z\sim 12$ to $\gtorder 0.3$ at $z\sim 6.3$. We show that the CR
galaxies can be easily detected by the current observations with the HST at $z\ltorder 10$, and some of the
most massive UCR galaxies can be detected for $z\ltorder 7$.
We find that the emerging UV flux from galaxies increases linearly with $\Mst$, but due to the mass dependence
of $\fesc$, the increase rate becomes much slower, i.e., $F_\nu^{\rm UV}\propto \Mst^{0.52}$.

\item The most massive CR galaxies have IR luminosities of $L_{\rm IR}\sim 6\times 10^{11}\,L_\odot$ at 
$z\sim 10$, and $\sim 4\times 10^{12}\,L_\odot$ at $z\sim 6.3$, and  $L_{\rm IR}\sim \Mst^{1.1}$. 
Massive galaxies appear bright at the sub-millimetre band because of the high SFR\,$\gtrsim 100\,\Msunyr$, 
and a strong dust extinction, $\fesc\lesssim 0.1$. The 1.1\,mm flux exhibits a weaker dependence on $\Mst$
than  $L_{\rm IR}$, i.e., $S_{1.1}\sim M^{0.9}$. This flux is $\gtrsim 10^{-2} \,{\rm mJy}$, and 
hence massive high-redshift galaxies can be detected by ALMA  up to $z\sim 10$. 
On the other hand, most of the UCR galaxies are difficult to detect by ALMA due to the lower SFR and dust amount. 
The fluxes of metal lines of C and O, and the CO molecular lines can be higher than the continuum emission from dust, 
and may enhance their detectability \citep{Inoue14}, although the physical state of the metals (e.g., their 
ionization levels and molecular fractions) are not clear. 
 
\item Some galaxies show disky shapes, which causes the large dispersion in the emerging UV flux 
depending on the viewing angle. Massive galaxies tend to be disky due to the efficient gas accretion from 
the IGM. 

\item The SFR, stellar mass, dust mass and other parameters of our massive CR galaxies agree broadly with the 
observed SMGs at $z\sim 3$ \citep[e.g.,][]{Hatsukade11}, 
and they correspond roughly to the bright SMG at $z = 6.3$, discovered by \citet{Riechers13}, which exhibits 
$\SFR\sim 2900\,\Msun$, $\Mst = 3.7\times 10^{10}\,\Msun$ and $\Mdust\sim 1.3\times 
10^9\,\Msun$. 

\end{itemize}

The observed properties of high-redshift galaxies have been used to determine their intrinsic properties and
to constrain the ongoing physical processes. One of the most important parameters appears to be the
escape fraction of the UV photons, both ionizing and non-ionizing. Here we focused on the latter one and determine
its effect on the measurable quantities, like the attenuated UV flux and the galaxy luminosity function.
This has been done comparing the normal and substantially overdense regions in the universe. We did not
find any substantial differences between the CR and UCR runs, except those which can be directly linked
to the environmental contrast, i.e., the background density. The associated accelerated growth of the
CR galaxies can indeed produce the massive quasar hosts by $z\sim 6$. Their LF appears to exceed that
of the UCR galaxies by a factor of 10--50 already at $z\ltorder 12$, and to extend to the absolute
magnitudes of $\sim -22$ by $z\sim 6$.
  
We have applied the SN dust model \citep{Todini01} and the constant dust-to-metal mass ratio based on the 
observations of local star-forming galaxies \citep{Draine07}.
On the other hand, dust properties of high-redshift galaxies are not understood yet, and there is 
still room for varying the dust size, composition, and the dust-to-metal mass ratio. 
Different dust models lead to distinct opacity curves as a function of wavelength, resulting in the 
different spectral properties \citep{Yajima14b}. 
However, if photons are mainly absorbed by the high-density clumps, $\fesc$ and emerging $\muv$ are 
not sensitive to the dust models. 
This ensues because most of the photons crossing these clumps are absorbed regardless of the dust properties 
within reasonable parameter ranges \citep{Yajima14b}. 
In this case, $\fesc$ and $\muv$ are determined mainly by the covering factor of the gas clumps. 
On the other hand, the IR properties of the emerging radiation can vary with the dust temperature dependence 
on the grain size \citep{Cen14}. 
At present, it is hard to constrain the dust temperature in galaxies at $z\gtrsim 6$ from the  
observational data, because of the difficulty to detect sub-millimetre fluxes at several frequencies. 

Recently, \citet{Kimm13} and \citet{Cen14} investigated UV and IR galaxy properties in
overdense region at $z=7$, by using cosmological adaptive mesh refinement (AMR) simulations and radiative 
transfer calculations. They have shown that  $M_{\rm star}\sim 5\times 10^8-2.5\times 10^{10}\, \Msun$ 
galaxies have been heavily dust obscured and their UV 
properties could be modified by the dust models. We have obtained similar results for $z=7$, however, our 
larger sample extending over a substantial range in redshift, as well as dramatically larger 
range of environment by considering the CR run, and therefore of the galaxy masses, allow to analyze the 
redshift evolution and detailed mass dependence of the observational 
properties of high-redshift galaxies, and the impact of galaxy morphology on their observational appearance.
In addition, although our simulations have shown strong extinction, the slope $\beta$ in 
$F\propto\lambda^{\beta}$ does not 
increase. This happens because of the presence of a bump at $\lambda\sim 2175\,{\rm \AA}$ 
\citep{Li08} in our dust model of type~II SN. Similar results have been obtained also by \citet{Kimm13}. 
A large parameter space in dust properties should be considered in the future studies and constrained
by observations. 

The overall mass range of galaxies in the UCR run and the lower mass galaxies in the CR run overlap with that 
in \citet{Yajima14c}. However, $\fesc$ of our simulations is somewhat lower in comparison. 
We have attributed this to differences in the SF recipes. 
The {\it Pressure} SF model \citep[e.g.,][]{DallaVecchia08, Schaye08, Choi10} have shown slower SFRs than the 
SF model based on the gas temperature and density used in \citet{Yajima14c}. 
The {\it Pressure} model allows the gas to evolve to high-density, $n_{\rm H} > 100\,{\rm cm^{-3}}$, by 
delaying the SF. In the presence of high-density gas clumps, absorption of stellar radiation becomes more 
efficient, leading to somewhat lower $\fesc$. We note that at present, no preferred SF recipe in numerical 
simulations exist. One hopes that via comparison and constraining with observations, the correct physics 
of SF will be developed. 

Massive galaxies with elevated SFR can be sources for the cosmic reionization. 
However, the mass dependence of $\fesc$ for ionizing photons can suppress the ionizing photon emissivity
of massive galaxies substantially \citep{Yajima11, Paardekooper13, Wise14}. 
In most cases, the escape fraction of the ionizing photons, $\fescion$, will be smaller than that of the UV 
continuum photons \citep{Yajima14c}. The former are also absorbed by hydrogen, whose optical depth can be 
much higher than that of dust. Therefore, $\fescion$ of massive galaxies can be much smaller than 0.1. 
Their rarity and small $\fescion$ make it doubtful that massive galaxies can serve as a major source of cosmic 
reionization.

The indicated upper limit of $\fescion$ for massive galaxies are smaller than that assumed in the observational 
studies \citep[e.g.,][]{Ouchi09b, Bouwens12}.
Wise et al. (2014) have shown recently that $\fescion$ decrease with the increasing stellar (galaxy) mass, and 
$\fescion < 0.1$ even for $M_{\rm star}\sim 10^6\,\Msun$ \citep[see also,][]{Yajima11, Yajima14c}. 
On the other hand, \citet{Kimm14} have shown that even galaxies with $\Mst\sim 10^{6} - 10^{8}\,\Msun$ 
can have $\fescion\gtrsim 0.1$.
Although the values of $\fescion$ are being debated, most of the theoretical works agree that $\fescion$ 
decreases with increasing galactic stellar mass. 
Hence, massive galaxies like the observed bright LBGs and SMGs at $z\gtrsim 6$ are not likely to have high $\fescion$. 
Some observations have shown that $\fescion\gtrsim 0.1$ \citep[e.g.,][]{Iwata09, Cooke14}, 
while others point to $\fescion \lesssim 0.05$ \citep[e.g.,][]{Siana10, Vanzella10}. 
On the other hand, low-mass faint galaxies can have higher $\fescion$ \citep{Yajima11, Wise14, Kimm14}, and serve as the main ionizing 
sources, if the faint-end slope of the LF is steep \citep[$\alpha \lesssim -2$, ][]{Bouwens12, Jaacks12}. 
Future telescopes, such as the JWST or 30\,m-class telescopes, will reveal the abundance of faint galaxies and 
allow to determine the ionizing sources in the high-redshift universe. 

In summary, we have analysed the observational appearance of high-redshift galaxies evolving in rare
overdense environment as well as in the average region of the universe, using the UV and IR bands and have
reproduced their LFs. The average field galaxy LF (UCR run) has been compared with the observationally-deduced 
LFs and have been found in a very reasonable agreement. The overall
impression is that these galaxies are compact, their SF sites appear to be heavily concentrated in the
central regions, and are enshrouded in dust produced by the SN. The centrally-peaked dust distribution,
therefore, results in substantial attenuating column densities and low escape fractions for the continuum UV 
photons, which decrease with the stellar mass in galaxies.
We find that the redshift evolution of $\fesc$ affects mostly the low-mass galaxies of 
$ \Mstar \ltorder 10^9\, \Msun$, where $\fesc$ increases by a factor of a few by 
$z\sim 6$. The emerging UV fluxes
from massive galaxies in the overdense fields nevertheless can be detected by the current HST observations, 
but the detection depends on the galaxy aspect angle, if indeed there is a tendency for the more massive
galaxies to be disky at these redshifts. Finally, we find that these massive galaxies above $\sim 
10^{9.5}\,M_{\rm star}$ can be detected by ALMA below $z\sim 10$ using reasonable integration time.

%
%
\section*{Acknowledgments}
We thank Volker Springel for providing us with the original version of GADGET-3,
and are grateful to Jun-Hwan Choi for valuable discussions. 
We thank the anonymous referee for useful comments.
This work has been supported under the International Joint Research Promotion Program by
Osaka University.  I.S. has benefited from the partial support by the NSF and STScI.
STScI is operated by AURA Inc., under NASA contract NAS 5-26555. 
E.R.D. thanks DFG  for support under SFB 956. 
K.N. acknowledges the partial support by JSPS KAKENHI Grant Number 26247022. 

%
%


\begin{thebibliography}{86}
\expandafter\ifx\csname natexlab\endcsname\relax\def\natexlab#1{#1}\fi

\bibitem[{{Biffi} \& {Maio}(2013)}]{Biffi13}
{Biffi} V., {Maio} U., 2013, \mnras, 436, 1621

\bibitem[{{Bouwens} {et~al.}(2007){Bouwens}, {Illingworth}, {Franx}, \&
  {Ford}}]{Bouwens07}
{Bouwens} R.~J., {Illingworth} G.~D., {Franx} M., {Ford} H., 2007, \apj, 670,
  928

\bibitem[{{Bouwens} {et~al.}(2009){Bouwens}, {Illingworth}, {Franx}, {Chary},
  {Meurer}, {Conselice}, {Ford}, {Giavalisco}, \& {van Dokkum}}]{Bouwens09}
{Bouwens} R.~J. et al., 2009,
  \apj, 705, 936

\bibitem[{{Bouwens} {et~al.}(2010){Bouwens}, {Illingworth}, {Oesch}, {Trenti},
  {Stiavelli}, {Carollo}, {Franx}, {van Dokkum}, {Labb{\'e}}, \&
  {Magee}}]{Bouwens10}
{Bouwens} R.~J. et al., 2010, \apjl, 708, L69

\bibitem[{{Blanc} {et~al.}(2011){Blanc}, {Adams}, {Gebhardt}, {Hill}, {Drory},
  {Hao}, {Bender}, {Ciardullo}, {Finkelstein}, {Fry}, {Gawiser}, {Gronwall},
  {Hopp}, {Jeong}, {Kelzenberg}, {Komatsu}, {MacQueen}, {Murphy}, {Roth},
  {Schneider}, \& {Tufts}}]{Blanc11}
{Blanc} G.~A. et al., 
  2011, \apj, 736, 31

\bibitem[{{Bouwens} {et~al.}(2011){Bouwens}, {Illingworth}, {Oesch},
  {Labb{\'e}}, {Trenti}, {van Dokkum}, {Franx}, {Stiavelli}, {Carollo},
  {Magee}, \& {Gonzalez}}]{Bouwens11b}
{Bouwens} R.~J. et al., 2011, \apj, 737, 90

\bibitem[{{Bouwens} {et~al.}(2012){Bouwens}, {Illingworth}, {Oesch}, {Trenti},
  {Labb{\'e}}, {Franx}, {Stiavelli}, {Carollo}, {van Dokkum}, \&
  {Magee}}]{Bouwens12}
{Bouwens} R.~J. et al.,
  2012, \apjl, 752, L5

\bibitem[{{Bouwens} {et~al.}(2014a)}]{Bouwens14a}
{Bouwens} R.~J. et al., 2014a, \apj, 793, 115

\bibitem[{{Bouwens} {et~al.}(2014b){Bouwens}, {Illingworth}, {Oesch}, {Trenti},
  {Labbe'}, {Bradley}, {Carollo}, {van Dokkum}, {Gonzalez}, {Holwerda},
  {Franx}, {Spitler}, {Smit}, \& {Magee}}]{Bouwens14}
{Bouwens} R.~J. et al., 2014b, arXiv:1403.4295

\bibitem[{{Calzetti} {et~al.}(2000)}]{Calzetti00}
{Calzetti} D., {Armus} L., {Bohlin} R.~C., {Kinney} A.~L., {Koornneef} J., 
{Storchi-Bergmann} T., 2000, \apj, 533, 682
	
\bibitem[{{Cen} \& {Zheng}(2013)}]{Cen12}
{Cen} R., {Zheng} Z., 2013, \apj, 775, 112

\bibitem[{{Cen} \& {Kimm}(2014)}]{Cen14}
{Cen} R., {Kimm} T., 2014, \apj, 782, 32

\bibitem[{{Choi} \& {Nagamine}(2009)}]{Choi09}
{Choi} J.-H., {Nagamine} K., 2009, MNRAS, 393, 1595

\bibitem[{{Choi} \& {Nagamine}(2010)}]{Choi10}
---, 2010, \mnras, 407, 1464

\bibitem[{{Choi} \& {Nagamine}(2011)}]{Choi11}
---, 2011, \mnras, 410, 2579

\bibitem[{{Cooke} {et~al.}(2014)}]{Cooke14}
{Cooke} J., {Ryan-Weber} E.~V., {Garel} T., {D{\'{\i}}az} C.~G., 2014, \mnras, 
441, 837

\bibitem[{{Dalla Vecchia} \& {Schaye}(2008)}]{DallaVecchia08}
{Dalla Vecchia} C., {Schaye} J., 2008, \mnras, 387, 1431

\bibitem[{{Dayal} \& {Ferrara}(2012)}]{Dayal12}
{Dayal} P., {Ferrara} A., 2012, \mnras, 421, 2568

\bibitem[{{Dayal} {et~al.}(2013){Dayal}, {Dunlop}, {Maio}, \&
  {Ciardi}}]{Dayal13}
{Dayal} P., {Dunlop} J.~S., {Maio} U., {Ciardi} B., 2013, \mnras, 434, 1486

\bibitem[{{Di Matteo} {et~al.}(2005){Di Matteo}, {Springel}, \&
  {Hernquist}}]{DiMatteo05}
{Di Matteo} T., {Springel} V., {Hernquist} L., 2005, \nat, 433, 604

\bibitem[{{Draine} (2003)}]{Draine03}
{Draine} B.~T., 2003, \araa, 41, 241

\bibitem[{{Draine} {et~al.}(2007)}]{Draine07}
{Draine} B.~T., {et~al.}, 2007, \apj, 663, 866

\bibitem[{{Dunkley} {et~al.}(2009)}]{Dunkley09}
{Dunkley} J., {et~al.}, 2009, \apj, 701, 1804

\bibitem[{{Dunlop} {et~al.}(2012)}]{Dunlop12}
{Dunlop} J.~S., {McLure} R.~J., {Robertson} B.~E., 
	{Ellis} R.~S., {Stark} D.~P.,  {Cirasuolo} M., {de Ravel} L., 2012, \mnras, 420, 901
	
\bibitem[{{Dunlop} {et~al.}(2013){Dunlop}, {Rogers}, {McLure}, {Ellis},
  {Robertson}, {Koekemoer}, {Dayal}, {Curtis-Lake}, {Wild}, {Charlot},
  {Bowler}, {Schenker}, {Ouchi}, {Ono}, {Cirasuolo}, {Furlanetto}, {Stark},
  {Targett}, \& {Schneider}}]{Dunlop13}
{Dunlop} J.~S. et al., 2013,
  \mnras, 432, 3520

\bibitem[{{Eisenstein} \& {Hut}(1998)}]{Eisenstein98}
{Eisenstein} D.~J., {Hut} P., 1998, \apj, 498, 137

\bibitem[{{Ellis} {et~al.}(2013){Ellis}, {McLure}, {Dunlop}, {Robertson},
  {Ono}, {Schenker}, {Koekemoer}, {Bowler}, {Ouchi}, {Rogers}, {Curtis-Lake},
  {Schneider}, {Charlot}, {Stark}, {Furlanetto}, \& {Cirasuolo}}]{Ellis13}
{Ellis} R.~S. et al., 2013, \apjl, 763, L7

\bibitem[{{Fan} {et~al.}(2003){Fan}, {etal.}}]{Fan.etal:03}
{Fan} X. et al., 2003, \aj, 125, 1649

\bibitem[{{Faucher-Gigu{\`e}re} {et~al.}(2009)}]{Faucher-Giguere09}
{Faucher-Gigu{\`e}re} C.-A., {Lidz} A., {Zaldarriaga} M., {Hernquist} L., 2009, \apj, 703, 1416
	
\bibitem[{{Finkelstein} {et~al.}(2013){Finkelstein}, {Papovich}, {Dickinson},
  {Song}, {Tilvi}, {Koekemoer}, {Finkelstein}, {Mobasher}, {Ferguson},
  {Giavalisco}, {Reddy}, {Ashby}, {Dekel}, {Fazio}, {Fontana}, {Grogin},
  {Huang}, {Kocevski}, {Rafelski}, {Weiner}, \& {Willner}}]{Finkelstein13}
{Finkelstein} S.~L. et al., 2013, \nat, 502, 524

\bibitem[{{Finkelstein} {et~al.}(2014)}]{Finkelstein14}
{Finkelstein} S.~L. et al., 2014, arXiv:1410.5439

\bibitem[{{Finlator} {et~al.}(2011){Finlator}, {Oppenheimer}, \&
  {Dav{\'e}}}]{Finlator11}
{Finlator} K., {Oppenheimer} B.~D., {Dav{\'e}} R., 2011, \mnras, 410, 1703

\bibitem[{{Gallerani} {et~al.}(2010){Gallerani}, {Maiolino}, {Juarez},
  {Nagao}, {Marconi}, {Bianchi}, {Schneider}, {Mannucci}, {Oliva}, {Willott},
  {Jiang}, \& {Fan}}]{Gallerani10}
{Gallerani} S. et al., 2010, A\&A, 523, 85


\bibitem[{{Gronwall} {et~al.}(2007){Gronwall}, {Ciardullo}, {Hickey},
  {Gawiser}, {Feldmeier}, {van Dokkum}, {Urry}, {Herrera}, {Lehmer}, {Infante},
  {Orsi}, {Marchesini}, {Blanc}, {Francke}, {Lira}, \& {Treister}}]{Gronwall07}
{Gronwall} C. et al., 2007, \apj, 667, 79

\bibitem[{{Hatsukade} {et~al.}(2011){Hatsukade}, {Kohno}, {Aretxaga},
  {Austermann}, {Ezawa}, {Hughes}, {Ikarashi}, {Iono}, {Kawabe}, {Khan},
  {Matsuo}, {Matsuura}, {Nakanishi}, {Oshima}, {Perera}, {Scott}, {Shirahata},
  {Takeuchi}, {Tamura}, {Tanaka}, {Tosaki}, {Wilson}, \& {Yun}}]{Hatsukade11}
{Hatsukade} B. et al., 2011, \mnras, 411, 102

\bibitem[{{Heller} {et~al.}(2007){Heller}, {Shlosman}, \&
  {Athanassoula}}]{Heller07}
{Heller} C.~H., {Shlosman} I., {Athanassoula} E., 2007, \apj, 671, 226

\bibitem[{{Hirashita} {et~al.}(2014){Hirashita}, {Ferrara}, {Dayal}, \&
  {Ouchi}}]{Hirashita14}
{Hirashita} H., {Ferrara} A., {Dayal} P., {Ouchi} M., 2014, \mnras, 443, 1704

\bibitem[{{Hinshaw} {et~al.}(2013)}]{Hinshaw13}
Hinschaw G. et al., 2013, \apjs, 208, 19
  
\bibitem[{{Hoffman} \& {Ribak}(1991)}]{Hoffman91}
{Hoffman} Y., {Ribak} E., 1991, \apjl, 380, L5

\bibitem[{{Hopkins} \& {Beacom}(2006)}]{Hopkins06SFR}
{Hopkins} A.~M., {Beacom} J.~F., 2006, \apj, 651, 142

\bibitem[{{Inoue} {et~al.}(2006){Inoue}, {Buat}, {Burgarella}, {Panuzzo},
  {Takeuchi}, \& {Iglesias-P{\'a}ramo}}]{Inoue06}
{Inoue} A.~K., {Buat} V., {Burgarella} D., {Panuzzo} P., {Takeuchi} T.~T.,
  {Iglesias-P{\'a}ramo} J., 2006, \mnras, 370, 380

\bibitem[{{Inoue} {et~al.}(2014){Inoue}, {Shimizu}, {Tamura}, {Matsuo},
  {Okamoto}, \& {Yoshida}}]{Inoue14}
{Inoue} A.~K., {Shimizu} I., {Tamura} Y., {Matsuo} H., {Okamoto} T., {Yoshida}
  N., 2014, \apjl, 780, L18

\bibitem[{{Iye} {et~al.}(2006){Iye}, {Ota}, {Kashikawa}, {Furusawa},
  {Hashimoto}, {Hattori}, {Matsuda}, {Morokuma}, {Ouchi}, \&
  {Shimasaku}}]{Iye06}
{Iye} M. et al., 2006, \nat, 443,
  186
\bibitem[{{Iwata} {et~al.}(2009)}]{Iwata09}
{Iwata} I. et al., 2009, \apj, 692, 1287

\bibitem[{{Jaacks} {et~al.}(2012){Jaacks}, {Choi}, {Nagamine}, {Thompson}, \&
  {Varghese}}]{Jaacks12}
{Jaacks} J., {Choi} J.-H., {Nagamine} K., {Thompson} R., {Varghese} S., 2012,
  \mnras, 420, 1606

\bibitem[{{Jaacks} {et~al.}(2013){Jaacks}, {Thompson}, \&
  {Nagamine}}]{Jaacks13}
{Jaacks} J., {Thompson} R., {Nagamine} K., 2013, \apj, 766, 94

\bibitem[{{Kennicutt}(1998a)}]{Kennicutt98}
{Kennicutt} Jr. R.~C., 1998, \araa, 36, 189

\bibitem[{{Kennicutt}(1998b)}]{Kennicutt98b}
{Kennicutt} Jr. R.~C., 1998, \apj, 498, 541

\bibitem[{{Kimm} \& {Cen}(2013)}]{Kimm13}
{Kimm} T., {Cen} R., 2013, \apj, 776, 35

\bibitem[{{Kimm} \& {Cen}(2014)}]{Kimm14}
{Kimm} T., {Cen} R., 2014, \apj, 788, 121

\bibitem[{{Leitherer} {et~al.}(1999){Leitherer}, {Schaerer}, {Goldader},
  {Delgado}, {Robert}, {Kune}, {de Mello}, {Devost}, \&
  {Heckman}}]{Leitherer99}
{Leitherer} C. et al., 1999,
  \apjs, 123, 3

\bibitem[{{Li} {et~al.}(2007){Li}, {Hernquist}, {Robertson}, {Cox}, {Hopkins},
  {Springel}, {Gao}, {Di Matteo}, {Zentner}, {Jenkins}, \& {Yoshida}}]{Li07}
{Li} Y. et al., 2007, \apj, 665, 187

\bibitem[{{Li} {et~al.}(2008){Li}, {Hopkins}, {Hernquist}, {Finkbeiner}, {Cox},
  {Springel}, {Jiang}, {Fan}, \& {Yoshida}}]{Li08}
{Li} Y. et al.,  2008, \apj, 678, 41

\bibitem[{{Madau} \& {Dickinson}(2014)}]{Madau14}
{Madau} P., {Dickinson} M., 2014, \araa, 52, 415

\bibitem[{{Madau} {et~al.}(1999){Madau}, {Haardt}, \& {Rees}}]{Madau99}
{Madau} P., {Haardt} F., {Rees} M.~J., 1999, ApJ, 514, 648

\bibitem[{{Maiolino} {et~al.}(2008)}]{Maiolino08}
{Maiolino} R. et al., 2008, A\&A, 488, 463

\bibitem[{{McLure} {et~al.}(2013){McLure}, {Dunlop}, {Bowler}, {Curtis-Lake},
  {Schenker}, {Ellis}, {Robertson}, {Koekemoer}, {Rogers}, {Ono}, {Ouchi},
  {Charlot}, {Wild}, {Stark}, {Furlanetto}, {Cirasuolo}, \&
  {Targett}}]{McLure13}
{McLure} R.~J. et al., 2013, \mnras, 432, 2696

\bibitem[{{Meurer} {et~al.}(1999)}]{Meurer99}
{Meurer} G.~R., {Heckman} T.~M., {Calzetti} D., 1999, \apj, 521, 64

\bibitem[{{Micha{\l}owski} {et~al.}(2012){Micha{\l}owski}, {Dunlop},
  {Cirasuolo}, {Hjorth}, {Hayward}, \& {Watson}}]{Michalowski12}
{Micha{\l}owski} M.~J., {Dunlop} J.~S., {Cirasuolo} M., {Hjorth} J., {Hayward}
  C.~C., {Watson} D., 2012, \aap, 541, A85

\bibitem[{{Nakajima} {et~al.}(2013)}]{Nakajima13}
{Nakajima} K., {Ouchi} M., {Shimasaku} K., {Hashimoto} T., {Ono} Y., {Lee} J.~C., 
  2013, \apj, 769, 3
	
\bibitem[{{Nagamine} {et~al.}(2010a)}]{Nagamine10}
{Nagamine} K., {Choi} J.-H., {Yajima} H., 2010, \apjl, 725, 219

\bibitem[{{Nagamine} {et~al.}(2010b){Nagamine}, {Ouchi}, {Springel}, \&
  {Hernquist}}]{Nagamine10b}
{Nagamine} K., {Ouchi} M., {Springel} V., {Hernquist} L., 2010, \pasj, 62, 1455

\bibitem[{{Nagamine} {et~al.}(2004){Nagamine}, {Springel}, {Hernquist}, \&
  {Machacek}}]{Nagamine04e}
{Nagamine} K., {Springel} V., {Hernquist} L., {Machacek} M., 2004, MNRAS, 350,
  385

\bibitem[{{Nozawa} {et~al.}(2006){Nozawa}, {Kozasa}, \& {Habe}}]{Nozawa06}
{Nozawa} T., {Kozasa} T., {Habe} A., 2006, \apj, 648, 435

\bibitem[{{Nozawa} {et~al.}(2007){Nozawa}, {Kozasa}, {Habe}, {Dwek}, {Umeda},
  {Tominaga}, {Maeda}, \& {Nomoto}}]{Nozawa07}
{Nozawa} T., {Kozasa} T., {Habe} A., {Dwek} E., {Umeda} H., {Tominaga} N.,
  {Maeda} K., {Nomoto} K., 2007, \apj, 666, 955

\bibitem[{{Oesch} {et~al.}(2012){Oesch}, {Bouwens}, {Illingworth}, {Gonzalez},
  {Trenti}, {van Dokkum}, {Franx}, {Labb{\'e}}, {Carollo}, \&
  {Magee}}]{Oesch12}
{Oesch} P.~A. et al., 2012, \apj, 759, 135

\bibitem[{{Oesch} {et~al.}(2013){Oesch}, {Bouwens}, {Illingworth}, {Labb{\'e}},
  {Franx}, {van Dokkum}, {Trenti}, {Stiavelli}, {Gonzalez}, \&
  {Magee}}]{Oesch13}
{Oesch} P.~A. et al., 2013, \apj, 773, 75

\bibitem[{{Oesch} {et~al.}(2014){Oesch}, {Bouwens}, {Illingworth}, {Labb{\'e}},
  {Smit}, {Franx}, {van Dokkum}, {Momcheva}, {Ashby}, {Fazio}, {Huang},
  {Willner}, {Gonzalez}, {Magee}, {Trenti}, {Brammer}, {Skelton}, \&
  {Spitler}}]{Oesch14}
{Oesch} P.~A. et al., 2014, \apj, 786, 108

\bibitem[{{Ono} {et~al.}(2012){Ono}, {Ouchi}, {Mobasher}, {Dickinson},
  {Penner}, {Shimasaku}, {Weiner}, {Kartaltepe}, {Nakajima}, {Nayyeri},
  {Stern}, {Kashikawa}, \& {Spinrad}}]{Ono12}
{Ono} Y. et al., 2012, \apj, 744, 83

\bibitem[{{Ono} {et~al.}(2010){Ono}, {Ouchi}, {Shimasaku}, {Akiyama}, {Dunlop},
  {Farrah}, {Lee}, {McLure}, {Okamura}, \& {Yoshida}}]{Ono10A}
{Ono} Y. et al.,  2010, \mnras, 402, 1580

\bibitem[{{Ouchi} {et~al.}(2009){Ouchi}, {Mobasher}, {Shimasaku}, {Ferguson},
  {Fall}, {Ono}, {Kashikawa}, {Morokuma}, {Nakajima}, {Okamura}, {Dickinson},
  {Giavalisco}, \& {Ohta}}]{Ouchi09b}
{Ouchi} M. et al., 2009, \apj, 706, 1136

\bibitem[{{Ouchi} {et~al.}(2004){Ouchi}, {Shimasaku}, {Okamura}, {Furusawa},
  {Kashikawa}, {Ota}, {Doi}, {Hamabe}, {Kimura}, {Komiyama}, {Miyazaki},
  {Miyazaki}, {Nakata}, {Sekiguchi}, {Yagi}, \& {Yasuda}}]{Ouchi04}
{Ouchi} M. et al., 2004,
  \apj, 611, 660
  
 \bibitem[{{Ouchi} {et~al.}(2004)}]{Ouchi04b}
{Ouchi} M. et al., 2004,
  \apj, 611, 685

\bibitem[{{Ouchi} {et~al.}(2010){Ouchi}, {Shimasaku}, {Furusawa}, {Saito},
  {Yoshida}, {Akiyama}, {Ono}, {Yamada}, {Ota}, {Kashikawa}, {Iye}, {Kodama},
  {Okamura}, {Simpson}, \& {Yoshida}}]{Ouchi10}
{Ouchi} M. et al., 2010, \apj, 723, 869

\bibitem[{{Paardekooper} {et~al.}(2013){Paardekooper}, {Khochfar}, \& {Dalla
  Vecchia}}]{Paardekooper13}
{Paardekooper} J.-P., {Khochfar} S., {Dalla Vecchia} C., 2013, \mnras, 429, L94

\bibitem[{{Pawlik} {et~al.}(2011){Pawlik}, {Milosavljevi{\'c}}, \&
  {Bromm}}]{Pawlik11}
{Pawlik} A.~H., {Milosavljevi{\'c}} M., {Bromm} V., 2011, \apj, 731, 54

\bibitem[{{Pawlik} {et~al.}(2013){Pawlik}, {Milosavljevi{\'c}}, \&
  {Bromm}}]{Pawlik13}
---, 2013, \apj, 767, 59


\bibitem[{{Pettini} {et~al.}(2001)}]{Pettini01}
{Pettini} M. et al., 2001, \apj, 554, 981

\bibitem[{{Riechers} {et~al.}(2013){Riechers}, {Bradford}, {Clements},
  {Dowell}, {P{\'e}rez-Fournon}, {Ivison}, {Bridge}, {Conley}, {Fu}, {Vieira},
  {Wardlow}, {Calanog}, {Cooray}, {Hurley}, {Neri}, {Kamenetzky}, {Aguirre},
  {Altieri}, {Arumugam}, {Benford}, {B{\'e}thermin}, {Bock}, {Burgarella},
  {Cabrera-Lavers}, {Chapman}, {Cox}, {Dunlop}, {Earle}, {Farrah}, {Ferrero},
  {Franceschini}, {Gavazzi}, {Glenn}, {Solares}, {Gurwell}, {Halpern},
  {Hatziminaoglou}, {Hyde}, {Ibar}, {Kov{\'a}cs}, {Krips}, {Lupu}, {Maloney},
  {Martinez-Navajas}, {Matsuhara}, {Murphy}, {Naylor}, {Nguyen}, {Oliver},
  {Omont}, {Page}, {Petitpas}, {Rangwala}, {Roseboom}, {Scott}, {Smith},
  {Staguhn}, {Streblyanska}, {Thomson}, {Valtchanov}, {Viero}, {Wang},
  {Zemcov}, \& {Zmuidzinas}}]{Riechers13}
{Riechers} D.~A. et al., 2013, \nat, 496, 329

\bibitem[{{Romano-D{\'{\i}}az} {et~al.}(2009){Romano-D{\'{\i}}az}, {Shlosman},
  {Heller}, \& {Hoffman}}]{Romano-Diaz09}
{Romano-D{\'{\i}}az} E., {Shlosman} I., {Heller} C., {Hoffman} Y., 2009, \apj,
  702, 1250

\bibitem[{{Romano-Diaz} {et~al.}(2011a){Romano-Diaz}, {Shlosman}, {Trenti}, \&
  {Hoffman}}]{Romano-Diaz11a}
{Romano-Diaz} E., {Shlosman} I., {Trenti} M., {Hoffman} Y., 2011a, \apj, 736, 66

\bibitem[{{Romano-D{\'{\i}}az} {et~al.}(2011b){Romano-D{\'{\i}}az}, {Choi},
  {Shlosman}, \& {Trenti}}]{Romano-Diaz11b}
{Romano-D{\'{\i}}az} E., {Choi} J.-H., {Shlosman} I., {Trenti} M., 2011b, \apjl,
  738, L19

\bibitem[{{Romano-D{\'{\i}}az} {et~al.}(2014){Romano-D{\'{\i}}az}, {Shlosman},
  {Choi}, \& {Sadoun}}]{Romano-Diaz14}
{Romano-D{\'{\i}}az} E., {Shlosman} I., {Choi} J.-H., {Sadoun} R., 2014, \apjl,
  790, L32
  
\bibitem[{{Sadoun} {et~al.}(in preparation){Sadoun}, {Shlosman},
  {Choi}, \& {Romano-D{\'{\i}}az}}]{Sadoun15}
{Sadoun} R., {Shlosman} I., {Choi} J.-H., {Romano-D{\'{\i}}az} E., 2015, 
  in preparation  
  
 \bibitem[{{Salpeter} (1955)}]{Salpeter55}
Salpeter E. E., 1955, \apj, 121, 161

 \bibitem[{{Schaerer} (2003)}]{Schaerer03}
{Schaerer} D., 2003, A\&A, 397, 527

\bibitem[{{Schaerer} \& {de Barros} (2009)}]{Schaerer09}
{Schaerer} D., {de Barros} S., 2009, A\&A, 502, 423

\bibitem[{{Schaye} \& {Dalla Vecchia}(2008)}]{Schaye08}
{Schaye} J., {Dalla Vecchia} C., 2008, MNRAS, 383, 1210

\bibitem[{{Schneider} {et~al.}(2004){Schneider}, {Ferrara}, \&
  {Salvaterra}}]{Schneider04}
{Schneider} R., {Ferrara} A., {Salvaterra} R., 2004, \mnras, 351, 1379

\bibitem[{{Shapley} {et~al.}(2003){Shapley}, {Steidel}, {Pettini}, \&
  {Adelberger}}]{Shapley03}
{Shapley} A.~E., {Steidel} C.~C., {Pettini} M., {Adelberger} K.~L., 2003, \apj,
  588, 65

\bibitem[{{Shibuya} {et~al.}(2012){Shibuya}, {Kashikawa}, {Ota}, {Iye},
  {Ouchi}, {Furusawa}, {Shimasaku}, \& {Hattori}}]{Shibuya12}
{Shibuya} T., {Kashikawa} N., {Ota} K., {Iye} M., {Ouchi} M., {Furusawa} H.,
  {Shimasaku} K., {Hattori} T., 2012, \apj, 752, 114

\bibitem[{{Shimizu} {et~al.}(2014){Shimizu}, {Inoue}, {Okamoto}, \&
  {Yoshida}}]{Shimizu14}
{Shimizu} I., {Inoue} A.~K., {Okamoto} T., {Yoshida} N., 2014, \mnras, 440, 731

\bibitem[{{Shlosman}(2013)}]{Shlosman13}
{Shlosman} I., 2013, in Secular Evolution of Galaxies, (eds.) J.Falcon-Barroso 
  \& J.H.Knapen, Cambridge, UK: Cambridge University Press, p.555

\bibitem[{{Siana} {et~al.}(2010)}]{Siana10}
{Siana} B. et al., 2010, \apj, 723, 241

\bibitem[{{Sijacki} {et~al.}(2009){Sijacki}, {Springel}, \&
  {Haehnelt}}]{Sijacki.etal:09}
{Sijacki} D., {Springel} V., {Haehnelt}, M.G., 2009, \mnras, 400, 100

\bibitem[{{Springel}(2005)}]{Springel05e}
{Springel} V., 2005, MNRAS, 364, 1105

\bibitem[{{Springel} \& {Hernquist}(2002)}]{Springel02}
{Springel} V., {Hernquist} L., 2002, MNRAS, 333, 649

\bibitem[{{Springel} \& {Hernquist}(2003{\natexlab{a}})}]{Springel03a}
---, 2003{\natexlab{a}}, MNRAS, 339, 289

\bibitem[{{Springel} \& {Hernquist}(2003{\natexlab{b}})}]{Springel03b}
---, 2003{\natexlab{b}}, MNRAS, 339, 312

\bibitem[{{Springel} {et~al.}(2005){Springel}}]{Springel.etal:05}
{Springel} V. et al., 2005, Nature, 435, 629

\bibitem[{{Springel} {et~al.}(2008){Springel}, {Wang}, {Vogelsberger},
  {Ludlow}, {Jenkins}, {Helmi}, {Navarro}, {Frenk}, \& {White}}]{Springel08a}
{Springel} V. et al., 2008, MNRAS, 391, 1685

\bibitem[{{Stark} {et~al.}(2009){Stark}, {Ellis}, {Bunker}, {Bundy}, {Targett},
  {Benson}, \& {Lacy}}]{Stark09}
{Stark} D.~P., {Ellis} R.~S., {Bunker} A., {Bundy} K., {Targett} T., {Benson}
  A., {Lacy} M., 2009, \apj, 697, 1493

\bibitem[{{Stark} {et~al.}(2013){Stark}, {Schenker}, {Ellis}, {Robertson},
  {McLure}, \& {Dunlop}}]{Stark13}
{Stark} D.~P., {Schenker} M.~A., {Ellis} R., {Robertson} B., {McLure} R.,
  {Dunlop} J., 2013, \apj, 763, 129

\bibitem[{Steidel {et~al.}(1999)Steidel, Adelberger, Giavalisco, Dickinson, \&
  Pettini}]{Steidel99}
Steidel C.~C., Adelberger K.~L., Giavalisco M., Dickinson M., Pettini M., 1999,
  ApJ, 519, 1

\bibitem[{{Tanaka} \& {Li}(2014)}]{Tanaka14}
{Tanaka} T.~L., {Li} M., 2014, \apj, 439, 1092

\bibitem[{{Todini} \& {Ferrara}(2001)}]{Todini01}
{Todini} P., {Ferrara} A., 2001, \mnras, 325, 726

\bibitem[Tomassetti et al.(2014)]{Tomassetti14} 
{Tomassetti} M., {Porciani} C., {Romano-D\'{\i}az} E., {Ludlow} A.~D., 
 {Papadopoulos} P.~P., 2014, \mnras, 445, L124

\bibitem[van de Weygaert \& Bertschinger(1996)]{vdW96} 
{van de Weygaert} R., {Bertschinger} E., 1996, \mnras, 281, 84 

\bibitem[{{Vanzella} {et~al.}(2010)}]{Vanzella10}
{Vanzella} E. et al, 2010, \apj,  725, 1011

\bibitem[{{Vanzella} {et~al.}(2011){Vanzella}, {Pentericci}, {Fontana},
  {Grazian}, {Castellano}, {Boutsia}, {Cristiani}, {Dickinson}, {Gallozzi},
  {Giallongo}, {Giavalisco}, {Maiolino}, {Moorwood}, {Paris}, \&
  {Santini}}]{Vanzella11}
{Vanzella} E. et al.,
  2011, \apjl, 730, L35

\bibitem[{{Verhamme} {et~al.}(2012){Verhamme}, {Dubois}, {Blaizot}, {Garel},
  {Bacon}, {Devriendt}, {Guiderdoni}, \& {Slyz}}]{Verhamme12}
{Verhamme} A., {Dubois} Y., {Blaizot} J., {Garel} T., {Bacon} R., {Devriendt}
  J., {Guiderdoni} B., {Slyz} A., 2012, \aap, 546, A111

\bibitem[{{Willott} {et~al.}(2010){Willott}, {et al.}}]{Willott.etal:10}
{Willott} C.~J., {et al.}, 2010, \aj, 139, 906

\bibitem[{{Wise} {et~al.}(2012){Wise}, {Turk}, {Norman}, \& {Abel}}]{Wise12a}
{Wise} J.~H., {Turk} M.~J., {Norman} M.~L., {Abel} T., 2012, \apj, 745, 50

\bibitem[{{Wise} {et~al.}(2014){Wise}, {Demchenko}, {Halicek}, {Norman},
  {Turk}, {Abel}, \& {Smith}}]{Wise14}
{Wise} J.~H., {Demchenko} V.~G., {Halicek} M.~T., {Norman} M.~L., {Turk} M.~J.,
  {Abel} T., {Smith} B.~D., 2014, \mnras, 442, 2560

\bibitem[{{Yajima} {et~al.}(2011){Yajima}, {Choi}, \& {Nagamine}}]{Yajima11}
{Yajima} H., {Choi} J.-H., {Nagamine} K., 2011, \mnras, 412, 411

\bibitem[{{Yajima} {et~al.}(2012{\natexlab{a}}){Yajima}, {Li}, {Zhu}, \&
  {Abel}}]{Yajima12b}
{Yajima} H., {Li} Y., {Zhu} Q., {Abel} T., 2012{\natexlab{a}}, \mnras, 424, 884

\bibitem[{{Yajima} {et~al.}(2012{\natexlab{b}}){Yajima}, {Li}, {Zhu}, \&
  {Abel}}]{Yajima12f}
{Yajima} H., {Li} Y., {Zhu} Q., {Abel} T., 2012{\natexlab{b}}, submitted to ApJ, arXiv:1211.0014

\bibitem[{{Yajima} {et~al.}(2012{\natexlab{c}}){Yajima}, {Li}, {Zhu}, {Abel},
  {Gronwall}, \& {Ciardullo}}]{Yajima12c}
{Yajima} H., {Li} Y., {Zhu} Q., {Abel} T., {Gronwall} C., {Ciardullo} R.,
  2012{\natexlab{c}}, \apj, 754, 118
  
 \bibitem[{{Yajima} {et~al.}(2012{\natexlab{d}})}]{Yajima12d}
  {Yajima} H., {Choi} J.-H., {Nagamine} K., 2012{\natexlab{d}}, \mnras, 427, 2889

\bibitem[{{Yajima} {et~al.}(2014{\natexlab{a}}){Yajima}, {Li}, {Zhu}, {Abel},
  {Gronwall}, \& {Ciardullo}}]{Yajima14c}
{Yajima} H., {Li} Y., {Zhu} Q., {Abel} T., {Gronwall} C., {Ciardullo} R., 2014{\natexlab{a}}, 
  \mnras, 440, 776

\bibitem[{{Yajima} {et~al.}(2014{\natexlab{b}}){Yajima}, {Nagamine},
  {Thompson}, \& {Choi}}]{Yajima14b}
{Yajima} H., {Nagamine} K., {Thompson} R., {Choi} J.-H., 2014{\natexlab{b}},
  \mnras, 439, 3073

\end{thebibliography}

\label{lastpage}

\end{document}